%% file: article.tex
\documentclass[11pt, a4paper]{article}
\usepackage{jheppub}
\usepackage[utf8]{inputenc}
\usepackage[T1]{fontenc}
\usepackage{amsfonts}
\usepackage{icomma}
\usepackage{hyperref}
\usepackage{titlesec}
\usepackage{lmodern}
\usepackage{units}
\usepackage{marvosym}
\usepackage[usenames,dvipsnames,svgnames,table]{xcolor}
\usepackage{multirow}
\usepackage{float}
\usepackage{tensor}
\usepackage{slashed}
\declareslashed{}{\mbox{\scriptsize /}}{.1}{0}{\varepsilon}
\newcommand{\sep}{\ensuremath{\slashed\varepsilon}}

\declareslashed{}{\mbox{\scriptsize /}}{.1}{0}{E}
\newcommand{\sE}{\ensuremath{\slashed E}}

\declareslashed{}{\mbox{\scriptsize /}}{.1}{0}{\mathcal E}
\newcommand{\smcE}{\ensuremath{\slashed {\mathcal E}}}

\newcommand{\beq}{\begin{equation}}
\newcommand{\eeq}{\end{equation}}

\newcommand{\beqa}{\begin{eqnarray}}
\newcommand{\eeqa}{\end{eqnarray}}

\newcommand{\by}{\begin{eqnarray}}
\newcommand{\ey}{\end{eqnarray}}
\newcommand{\al}{\alpha}
\newcommand{\be}{\beta}
\newcommand{\ga}{\gamma}

\newcommand{\de}{\delta}
\newcommand{\ep}{\epsilon}
\newcommand{\om}{\omega}
\newcommand{\si}{\sigma}

\newcommand{\thalf}{\tfrac{1}{2}}

\newcommand{\pd}{\partial}

\newcommand{\Lam}{\Lambda}
\newcommand{\tM}{\tilde M}
\newcommand{\tP}{\tilde P}
\newcommand{\tK}{\tilde K}
\newcommand{\tD}{\tilde D}
\newcommand{\te}{\tilde e}
\newcommand{\tf}{\tilde f}
\newcommand{\tom}{\tilde \om}
\newcommand{\tb}{\tilde b}
\newcommand{\tLam}{\tilde \Lambda}

\newcommand{\mcO}{\mathcal O}
\newcommand{\mchO}{\mathcal {\hat O}}
\newcommand{\mcD}{\mathcal D}
\newcommand{\mchD}{\mathcal {\hat D}}

\newcommand{\con}[1]{\textcolor{ForestGreen}{\underline{#1}}}
\newcommand{\irr}[1]{{#1}}
\newcommand{\repcontent}[1]{|_{\pmb{#1}}}
\newcommand{\mr}[1]{\mathrm{#1}}
\newcommand{\comment}[1]{}
\newcommand{\transl}[1]{$\textsc{transl}_#1$}
\newcommand{\scale}[1]{$\textsc{scale}_#1$}
\newcommand{\lorentz}[1]{$\textsc{Lorentz}_#1$}
\newcommand{\trace}{\mr{trace}}

\title{The non-linear coupled spin 2 - spin 3 Cotton equation in three dimensions}
\author[a]{Hampus Linander,} 
\author[a]{Bengt E.W. Nilsson}
\affiliation[a]{
Department of Physics\\
Theoretical Physics\\
Chalmers University of Technology\\
S-412 96 Göteborg, Sweden
}
\emailAdd{linander@chalmers.se}
\emailAdd{tfebn@chalmers.se}
\abstract{In the context of three-dimensional conformal higher spin theory  we derive, in the frame field formulation, the full non-linear spin 3 Cotton equation coupled to spin 2. This is done by solving the corresponding Chern-Simons gauge theory system of equations, that is, using $F=0$ to eliminate all auxiliary fields and thus  expressing the Cotton equation in terms of just the spin 3 frame field and spin 2 covariant derivatives and tensors (Schouten). In this derivation we
neglect the spin 4 and higher spin sectors and approximate the star product commutator by   a Poisson bracket. The resulting spin 3 Cotton equation is complicated but can be related to linearized versions in the metric formulation obtained previously by other authors.
The expected symmetry (spin 3 ``translation'', ``Lorentz'' and ``dilatation'') properties are verified for Cotton and other relevant tensors but  some perhaps unexpected features emerge in the process, in particular in relation to the non-linear equations.
We discuss the structure of this non-linear spin 3 Cotton equation but its explicit form is only presented here, in an exact but not completely refined version, in appended files obtained by computer algebra methods.
Both the frame field and metric formulations are  provided.}


\begin{document}
\maketitle
\flushbottom
\section{Introduction}

Higher spin theories, see, e.g.,  \cite{Vasiliev:1999ba,Vasiliev:2003ev} and references therein, are of great interest because they purportedly  arise in the zero tension limit  of string theory and for the special  role they play in many examples of AdS/CFT \cite{Sundborg:2000wp, Wittenjhs60,Mikhailov:2002bp, Sezgin:2002rt}; for $AdS_4/CFT_3$ dualities involving vector models, see also \cite{Klebanov:2002ja, Leigh:2003gk, Sezgin:2003pt,Giombi:2009wh} and for $AdS_3/CFT_2$ see, e.g.,  \cite{Gaberdiel:2014cha, Gaberdiel:2015wpo}. However, they are also interesting for intrinsic reasons. Some of their most intriguing features stem from the fact that they describe  gauge fields of all (even) spins and for the huge challenge it has been to formulate a consistent interacting theory for such higher spin (HS) theories. Unfortunately, the solution to this problem  has turned out to be rather complicated.

The main method of construction is due to Vasiliev \cite{Vasiliev:1999ba, Vasiliev:2003ev} and works only in AdS and related backgrounds. In space-time dimensions $D\geq 4$  this method  
leads to rather intricate interaction 
 terms in the field equations (Vasiliev's equations) which  have so far been constructed explicitly
only for terms quadratic in the fields \cite{Sezgin:2002ru,  Boulanger:2015ova} (see also \cite{Boulanger:2008tg, Giombi:2009wh} for some specific aspects of such terms). One problem is  that there is
a proliferation of derivatives in the interaction terms, see, e.g.,  \cite{Metsaev:2005ar, Buchbinder:2006eq, Bekaert:2010hw} and the more recent work in 
\cite{ Kessel:2015kna,Boulanger:2015ova, Bekaert:2015tva, Skvortsov:2015lja}
 and references therein, while, of course, the kinetic terms are  quadratic in derivatives. In general  cubic vertices (in a Lagrangian) can be defined to contain at most $s_1+s_2+s_3$ derivatives (where $s_i$ are the three spins of the fields in the vertex) since terms with a higher number of derivatives can be removed by field redefinitions. For four and higher point vertices the phenomenon of derivative dressing, if it occurs as part of the Vasiliev equations, will most likely lead to an unbounded number of derivatives which can no longer be redefined away. Some recent papers have addressed the nature
of the potential non-localities  that may arise from interaction terms with an unlimited number of derivatives and suggested recipes  for the elimination of non-local
effects that could endanger the consistency of these theories. One approach to actually compute (or define) these higher point vertices in the Lagrangian is to deduce them from the
CFT at the boundary of AdS. This may rescue the situation as argued,  and also demonstrated, for  three and four point vertices in \cite{Bekaert:2015tva}. For Vasiliev theories which are  dual to {\it free} CFT, where any $n$-point function
can in principle be computed, this approach will hopefully provide a basis for an existence proof of the bulk higher spin theory  including a proper definition of the problematic vertices in AdS.

Another conceivable approach to getting a handle on the complicated structure of higher derivative 
terms that dress up any  given basic $n$-point vertex (defined to have the minimal number of derivatives possible) may be to start
with a conformal HS theory in the same dimension as the AdS HS theory. The complications due to derivative dressing does not exist in the conformal case since there is no dimensionful parameter. Then, provided one can find a conformal HS theory that in an AdS background gives rise to a Vasiliev type HS theory, one could perhaps gain a better understanding of the structure of  the multi-derivative terms and maybe even the intricacies involved in the construction of the Vasiliev equations and finding a Lagrangian formulation. 

A different  reason for studying three-dimensional conformal theories containing gauge fields of all spins $s=2,3,...$ was indicated by Giombi et al. in \cite{Giombi:2013yva}. There the authors wanted to compute the free energy in the Vasiliev $\mr{AdS}_4$ and compare it to the corresponding result in the boundary CFT$_3$. A relatively easy way to perform  this AdS/CFT check is to compute the free energies with the two possible boundary conditions and then subtract the results from each other. In order to make sense of this calculation one needs to argue that dynamics on the boundary arises for all spins when tuning the boundary conditions for the corresponding higher spin gauge fields in the bulk from Dirichlet to Neumann.\footnote{See also \cite{Vasiliev:2012vf} for comments concerning the possibility to use unconventional boundary conditions for bulk gauge fields of spin $\geq 3$ in AdS HS theory related to interacting HS Chern-Simons theories in the boundary. Similar issues  are also  discussed in earlier work, e.g., in \cite{Leigh:2003gk}.} At the parity invariant $A$ and $B$ points of the Vasiliev theory the dynamical CFT$_3$ on the boundary involve parity invariant induced actions while in between these $A$ and $B$ points  parity is broken which is related to the  presence of a spin one Chern-Simons term  on the boundary. In this situation probably also gauge fields in the bulk with spins two and higher can be assigned mixed boundary conditions again leading to Chern-Simons-like kinetic
terms for all higher spins on the boundary,
see, e.g., \cite{Giombi:2011kc, Chang:2012kt,Giombi:2012ms} and references therein\footnote{In fact, the spin one boundary condition may be tied to the  spin two and higher ones in highly supersymmetric cases as indicated by the topologically gauged $CFT_3$ constructed in \cite{Gran:2008qx, Chu:2009gi, Gran:2012mg} which connect the spin one and two Chern-Simons terms.}. For spin one the Chern-Simons phenomenon was discussed by Witten in \cite{Witten:2003ya}  whose arguments were subsequently used in the case of spin two in \cite{Leigh:2003ez, deHaro:2008gp}. These boundary CFT$_3$ theories with Chern-Simons terms are probably of the same kind as the ones we study in this paper. 
The first investigations of their non-linear structure and interactions with scalar fields were conducted 
in \cite{Nilsson2013, Nilsson2015}. The present work is a direct continuation of these
latter papers. 

Historically, pure (i.e., without matter fields\footnote{Spins $s<2$.}) three-dimensional conformal HS theory was analyzed at the linear level in \cite{Pope-Townsend89} where the representation theory needed for the elimination of the auxiliary fields was explained. At the non-linear level, in \cite{Fradkin:1989xt}  the authors  expanded (super)Chern-Simons theory  in terms of the HS component fields and computed the cubic interaction terms with their explicit coefficients from the star product. Note, however, that in this last work the auxiliary fields (both Stückelberg and dependent ones) were retained hiding all four and higher point vertices as well as  making, e.g.,  a comparison to the metric formulation impossible. This latter aspect is discussed in some detail for spin 3 in  later sections of this paper. One of the main goals of this paper is  to perform the elimination of the auxiliary fields in full detail and derive the resulting spin 3 Cotton equation.

In $D=2+1$ HS theories are very special. Here the infinite tower of spin states can be truncated down to some finite maximum value of the spin\footnote{If the Vasiliev construction is based on the HS algebra $hs(\mu)$ \cite{Prokushkin:1998bq, Vasiliev:2012vf}  all spins are in general required for consistency. If, however, $\mu=N$, an integer, the  generators for spins  $s\geq N+1$ form an ideal and the corresponding fields can  be truncated away \cite{Vasiliev:1989re}, see also, e.g., \cite{Gaberdiel:2012uj}.}. The resulting theories are Chern-Simons gauge theories based
on a finite dimensional HS algebra $\mr{sl}(N,\mathbb{R})\oplus \mr{sl}(N,\mathbb{R})$, generalizing  the Lie algebra  of the $\mr{AdS}_3$ isometry group   $\mr{sl}(2,\mathbb{R})\oplus \mr{sl}(2,\mathbb{R})$, and contain all spins from $2$ to $N$.
These theories  are rather easily written out in full detail (at least for small $N\geq 2$)  and many of their intriguing properties have been discussed in the literature, see, e.g., \cite{Ammon:2011nk, Fredenhagen:2014oua}.

In this paper we continue, in the spirit of \cite{Nilsson2013}, the study of conformal HS theories in $2+1$ dimensions. The HS algebra
is then related to the conformal algebra appropriate for $2+1$ dimensions, namely $\mr{so}(3,2)$,
which means that, as in higher dimensions, fields with all spins $s=2,3,4,..$ (or just the even ones) must be considered together. However, these theories are in some sense much simpler than the ones based on the  Vasiliev
construction of HS theories in higher dimensional $\mr{AdS}_{D\geq 4}$. As mentioned above, one reason for this is that
the infinite tail of higher derivative terms that can be added to any given current in AdS
does not exist in conformal HS theories since there is no dimensionful parameter that can be used to compensate the dimension of the extra derivatives. 

Higher order  derivative terms
appear also in  conformal HS theories but then the number of derivatives is unique given the field content of the kinetic or interaction term in question. E.g., in $D=2+1$ dimensions a spin $s$ kinetic term is of order $2s-1$ in derivatives. Note, however, that  in any given spin $s$ field equation  terms with an unlimited number of derivatives will always appear but  only in terms where the sum of the spins of the fields grows beyond all limits. These features become clear if one considers the
Chern-Simons version of the theory where the fundamental spin $s$ frame field has dimension $L^{s-2}$. If one Taylor expands the star product in the Chern-Simons HS gauge theory Lagrangian, as done in \cite{Fradkin:1989xt}, there is of course just one derivative appearing in the answer (in the kinetic terms and non in the cubic terms). But  the theory then contains  huge numbers of auxiliary fields, both dependent and Stückelberg ones. The multi-derivative properties of the theory then arise as a consequence of  eliminating these extra fields  \cite{Pope-Townsend89, Nilsson2013} but the key point is that the derivatives will now appear in the theory in a controlled way. As mentioned above, this leads to spin $s$ kinetic terms with $2s-1$ derivatives. As another example, consider an $A^3$ term from the original Chern-Simons Lagrangian. It will contain three spin $s_i$ dimensionless one-form cascade\footnote{The highest rank fields at each spin level, to be defined more precisely in section \ref{sec:spin3}.} fields $\tom_{s_i}(s_i-1,s_i-1)$ from the expansion\footnote{For three examples see eqs. (\ref{gaugefieldspintwo}), (\ref{gaugefieldspinthree}) and (\ref{gaugefieldspinfour}).} of the gauge fields $A_{s_i}$ (such that the tensor product of the three spins contains a singlet) and hence no derivatives. However, after elimination of the auxiliary fields, $\tom_{s_i}$ is expressed in terms of $s_i-1$ derivatives acting on the spin $s_i$ frame field. The number of derivatives in such a cubic term is then $s_1+s_2+s_3-3$. In the process of deriving these terms there will also appear terms, coming from commutators of covariant derivatives,  with more than three frame fields but the same number $s_1+s_2+s_3-3$ of derivatives\footnote{The reason for this is that we use exclusively  derivatives which are spin 2 Lorentz covariant and hence a commutator generates a spin 2 Ricci tensor which also has two derivatives.}. 

It is the purpose of this paper to continue the study in  \cite{ Nilsson2013, Nilsson2015} of the effects of eliminating the extra fields and in particular to derive the complete non-linear Cotton equation for the spin 3 frame field coupled to that of spin 2, the usual dreibein. 
These results have been obtained by means of a computer algebra system\footnote{A Mathematica based tensor algebra system developed for this purpose, using xPerm \cite{xperm} for tensor canonicalization.} and their extensive nature (over 1000 terms in the frame formulation) makes them unsuitable for presentation in the body of this text. The non-linear Cotton equation in the frame field and metric formulation is instead included in the appended files \cite{LinanderMath,LinanderMath2} in the form of unrefined output (see section \ref{sec:nonlinCotton3} for details).
We hope to give it in a more useful form elsewhere \cite{LinanderNilsson}. 

Finally, a somewhat different motivation for this work comes from the intriguing electric-magnetic duality properties discussed in \cite{Henneaux:2015cda} and the role played by  higher spin Cotton tensors in that context.

This paper is organized as follows. In section two we give some basic formulae explaining the structure of the HS algebra, the generators, the gauge field  and the Chern-Simons theory.
We also discuss the truncation of the full HS theory used in the rest of this paper and give a review of the spin 2 case. In section three we turn to the main subject of the paper namely the spin 3/spin 2 subsystem. There we introduce the gauge choice and discuss the cascade equations that finally makes it possible to obtain the spin three Cotton equation whose structure is 
 explained in section four. In that section also the linearized version and its relation to the metric formulation is clarified. Section 5 contains a first analysis of the cascade structure of the
linearized spin four sector thereby paving the way for a more thorough investigation of this sector in the future. Some conclusions are collected in section 6. Conventions and other useful information can be found in the appendix.
\section{Preliminaries}

The three-dimensional conformal group is $\mathrm{SO}(3,2)$ and its Lie algebra consists of the generators corresponding to  translations, $P_a$, Lorentz transformations, $M_a$, dilatations, $D$, together with the special conformal transformations, $K_a$. The Lorentz subalgebra is 
 $\mr{sl}(2,\mathbb{R})$ whose generators can be chosen as
 (see appendix \ref{app:A} for index conventions etc)
\begin{equation}
\label{matrixsltwo}
  T_{11} = \begin{pmatrix} 0 & 1 \\ 0 & 0 \end{pmatrix}, \quad T_{12} = \begin{pmatrix} -\tfrac{1}{2} & 0 \\ 0 & \tfrac{1}{2} \end{pmatrix}, \quad T_{22} = \begin{pmatrix} 0 & 0 \\ 1 & 0 \end{pmatrix},
\end{equation}
which can be compactly expressed as $\left( T_{\alpha \beta} \right)^\gamma{}_\delta=\delta^\gamma_{(\alpha}\epsilon_{\beta)\delta}$.

For the purpose of extending this to higher spins a convenient realization of these algebras is in terms of operators which are bilinear in a pair of $\mathrm{Spin}(2,1)\cong\mathrm{SL}(2,\mathbb{R})$ spinor variables $q^{\al}$ and $p_{\al}$ as will be described below. The generalization to higher spins is well-known and can be found in, e.g., \cite{Fradkin:1989xt, Pope-Townsend89}. Functions (of the symbols) of these operators are then multiplied by means of a star product. In this paper, however, we are mostly concerned with the classical approximation of the star product where
$q^{\al}$ and $p_{\al}$ are classical phase space variables.
Consequently, we will use a Poisson bracket instead of the star product commutator. 

\subsection{\texorpdfstring{Poisson bracket  realization of $\mr{so}(3,2)$}{Poisson bracket}}

By multiplying the above generators $\left( T_{\alpha \beta} \right)^\gamma{}_\delta=\delta^\gamma_{(\alpha}\epsilon_{\beta)\delta}$ by  coordinates $q_\ga$ and their conjugate momenta $p^{\de}$ we get 
\begin{equation}
  T_{\alpha \beta} = q_{(\alpha}p_{\beta)},
\end{equation}
where we used the fact that the spinor indices  on $q^{\al}$ and $p_{\al}$ can be raised and  lowered from the left by $\ep^{\al\be}$ and its inverse $\ep_{\al\be}$, respectively. 
Using the Poisson bracket
\begin{equation}
  \left\{f,g\right\}_{PB} = \tfrac{\partial f}{\partial q^\alpha}\tfrac{\partial g}{\partial p_\alpha} - \tfrac{\partial g}{\partial q^\alpha}\tfrac{\partial f}{\partial p_\alpha},
  \label{eq:poisson}
\end{equation}
one finds the same commutation relations as obtained using the matrix realization in (\ref{matrixsltwo}).
With the help of the three-dimensional gamma matrices\footnote{A convenient choice of real matrices $(\ga^a)_{\al}{}^{\be}$ is $\gamma^0 = i \sigma^2:=\ep$, $\gamma^1 = \sigma^1$, $\gamma^2 = \sigma^3$. For further conventions see appendix \ref{app:A}.} $\gamma^a$ the generators $T_{\al\be}$  correspond in the vector representation to
\begin{equation}
  M^a = -\tfrac{1}{2}\left(\gamma^a\right)_\alpha{}^\beta(q^\alpha p_\beta),
\end{equation}
which satisfy the familiar commutation relations ($\ep^{012}=1$ and $\ga^{abc}=\ep^{abc}{\bf 1}$)
\begin{equation}
  \left\{M^a,\,M^b\right\}_{\mathrm{PB}} = \epsilon^{a b}{}_c M^c.
\end{equation}
The rest of the generators for the conformal algebra $\mr{so}(3,2)$ are then realised as follows\footnote{Our convention is $q\cdot p = q^\alpha p_\alpha$. The spinor indices  
on the $\ga$-matrices are raised and lowered from the left  for the first index while for  the second it is done from the right. Thus, e.g., $(\ga^a)_{\al}{}^{\be}p_{\be}=(\ga^a)_{\al\be}p^{\be}$ which also defines $\gamma$-matrices with both indices down and similarly for two upper ones using $q^{\al}(\ga^a)_{\al}{}^{\be}=-q_{\al}(\ga^a)^{\al\be}$.}
\begin{equation}
  D=-\tfrac{1}{2}q\cdot p, \quad P^a = -\tfrac{1}{2}\left(\gamma^a\right)_{\alpha \beta}q^\alpha q^\beta, \quad K^a = -\tfrac{1}{2}\left(\gamma^a\right)^{\alpha\beta}p_\alpha p_\beta.
\end{equation}
Together they satisfy the commutation relations of the conformal algebra, where the remaining non-zero Poisson brackets are given by
\begin{equation}
  \left\{ M^a,\, P^b\right\}_{PB} = \epsilon^{a b}{}_c P^c, \quad \left\{ M^a,\, K^b\right\}_{PB} = \epsilon^{a b}{}_c K^c,
\end{equation}
and
\begin{equation}
  \left\{ D,\, P^a\right\}_{PB} = P^a, \quad \left\{ D,\, K^a\right\}_{PB} = -K^a, \quad \left\{ P^a,\, K^b\right\}_{PB} = -2\epsilon^{a b}{}_c M^c - 2\eta^{a b} D.
\end{equation}
These operators belong to the spin $j=1$ sector of the HS algebra while, as will be clear below, the corresponding gauge fields are part of the spin $s=2$ sector, i.e., the ordinary conformal gravity sector.

There is now a natural extension of this algebra to all integer spins where the generators are taken to be general even degree polynomials in $q^\alpha$ and $p_\beta$. One of the virtues of this realization is that the  generators in irreducible representations of the Lorentz group are easily  written down and the HS algebra rather straightforwardly computed, not only classically in terms of the Poisson bracket \cite{Pope-Townsend89} but also quantum mechanically in terms of multi-commutators or from expanding the star product commutator \cite{Fradkin:1989xt}. In this paper we will only use the classical variables which corresponds to a single commutator approximation of the star product.

\subsection{\texorpdfstring{The $\mr{so}(3,2)$ higher spin algebra}{The conformal higher spin algebra}}
In terms of the real commuting $\mathrm{SL}(2,\mathbb{R})$ spinors $q_\alpha$ and $p^\alpha$ the generators of the higher spin algebra are given by
\begin{equation}
  G(n_q,n_p,c)^{\alpha_1\ldots \alpha_{n_q+n_p-2c}} = \left(-\tfrac{1}{2}\right)^{\tfrac{n_q+n_p}{2}}q^{(\alpha_1}\cdots q^{\alpha_{n_q-c}} p^{\alpha_{n_q-c+1}}\cdots p^{\alpha_{n_p+n_q-2c})} (q\cdot p)^c, 
\end{equation}
where $n_q+n_p$ is a non-negative even integer for the HS theory we are interested in here.
Note that a generator is uniquely specified by the number of p's and q's present and how many of them are contracted. Since the only non-zero scalar product is  $q\cdot p$   we have that \mbox{$c \leq \mathrm{min}(n_q,n_p)$}.  Note that the generators are totally traceless due to the antisymmetry of the $\mathrm{SL}(2,\mathbb{R})$ metric $\epsilon_{a b}$ and thus belong to irreducible representations of $\mr{sl}(2,\mathbb{R})$.

The tensor presentation of these generators is then
\begin{multline}
  G(n_q,n_p,c)^{a_1 \ldots a_{N}} = (-1)^{\left\lfloor \tfrac{n_p-c}{2} \right\rfloor}\left(-\tfrac{1}{2}\right)^{\tfrac{n_q+n_p}{2}} \times \\ \times (\gamma^{a_1})_{\alpha_1 \alpha_2}\cdots (\gamma^{a_N})_{\alpha_{2N-1} \alpha_{2N} } q^{(\alpha_1}\cdots q^{\alpha_{n_q-c}} p^{\alpha_{n_q-c+1}}\cdots p^{\alpha_{2N})} (q\cdot p)^c,
\end{multline}
where $N=\tfrac{n_q+n_p}{2}-c$ and the vector indices $a_1...a_N$ are in the same irrep (that is symmetric and traceless which means that the label $c$ is redundant) as the $2N$ spinor indices.

For spin 2 the generators are organized as
\begin{equation}
  \begin{array}{l c l c l}
    G^a(2,0) &=& P^a &=& -\tfrac{1}{2}(\gamma^a)_{\alpha \beta}q^\alpha q^\beta,\\
    G^a(1,1) &=& M^a &=& -\tfrac{1}{2}(\gamma^a)_{\alpha}{}^{\beta}q^\alpha p_\beta,\\
    G(1,1) &=& D &=& -\tfrac{1}{2}q\cdot p,\\
    G^a(0,2) &=& K^a &=& -\tfrac{1}{2}(\gamma^a)^{\alpha\beta}p_\alpha p_\alpha,
\end{array}
\end{equation}
and generate, as already mentioned, the algebra of $\mathrm{SO}(3,2)$, the conformal group in three dimensions.

The spin 3 generators are
\begin{equation}
\label{spinthreegenerators}
  \begin{array}{l c l c l}
    G^{a b}(4,0) &=& P^{a b} &=& \tfrac{1}{4}(\gamma^a)_{\alpha\beta}(\gamma^b)_{\gamma\delta}q^\alpha q^\beta q^\gamma q^\delta, \\
    G^{a b}(3,1) &=& \tilde{P}^{a b} &=& \tfrac{1}{4}(\gamma^a)_{\alpha\beta}(\gamma^b)_{\gamma\delta}q^{(\alpha} q^\beta q^{\gamma}p^{\delta)}, \\
    G^a(3,1) &=& \tilde{P}^{a} &=& \tfrac{1}{4}(\gamma^a)_{\alpha \beta}q^\alpha q^\beta (q\cdot p),\\
    G^{a b}(2,2) &=& \tilde{M}^{a b} &=& -\tfrac{1}{4}(\gamma^a)_{\alpha\beta}(\gamma^b)_{\gamma\delta} q^{(\alpha} q^\beta p^\gamma p^{\delta)},\\
    G^a(2,2) &=& \tilde{M}^{a} &=& \tfrac{1}{4}(\gamma^a)_{\alpha \beta}q^\alpha p^\beta (q\cdot p),\\
    G(2,2) &=& \tilde{D} &=& \tfrac{1}{4}(q\cdot p)^2, \\
    G^{a b}(1,3) &=& \tilde{K}^{a b} &=& -\tfrac{1}{4}(\gamma^a)_{\alpha\beta}(\gamma^b)_{\gamma\delta}q^{(\alpha} p^\beta p^\gamma p^{\delta)},\\
    G^a(1,3) &=& \tilde{K}^{a} &=& -\tfrac{1}{4}(\gamma^a)_{\alpha \beta}p^\alpha p^\beta (q\cdot p),\\
    G^{a b}(0,4) &=& \tilde{K}^{a b} &=& \tfrac{1}{4}(\gamma^a)_{\alpha\beta}(\gamma^b)_{\gamma\delta}p^{\alpha} p^\beta p^\gamma p^{\delta},
  \end{array}
\end{equation}
where the minus signs come from raising the indices on all  the $p_{\al}$ spinors.
The higher spin algebra commutation relations can now easily be computed using the Poisson bracket. The result is gathered in appendix \ref{sec:hsalgebra}.

Before we continue with the analysis of the system for specific values of the spin (spin 4 is the subject of section 5) we should
explain what kind of approximations/truncations we are implementing. The  Chern-Simons theory used in this paper is based on the HS algebra expressed in terms of generators that are Weyl ordered even polynomials of the {\it operators} $q^{\al},\,p_{\al}$. One can then either compute the commutators of these or consider their symbols (where $q^{\al},\,p_{\al}$ are classical numbers) and use the Moyal star product commutator associated with ordinary  even polynomial functions of the classical $q^{\al},p_{\al}$. 
Letting $j\geq 1$ ($j=0$ appears only on the RHS of the second commutator below) denote the spin\footnote{Note that by spin $j$ is here related to the number of $q^{\al}$s and $p_{\al}$s by $j=\thalf(n_q+n_p)$ although the actual irrep associated with some of the generators have some lower spin value (if factors of $q^{\al}p_{\al}$ occur).}
of the generators $G(j)$, the structure of the star product commutators are
\beq
[G(j),G(j')]_*=G(j+j'-1)+G(j+j'-3)+...+G(1),
\eeq
if $j$ and $j'$ are both either even or odd, while
\beq
\label{mixedjcommutator}
[G(j),G(j')]_*=G(j+j'-1)+G(j+j'-3)+...+G(2),
\eeq
if one of  $j$ and $j'$ is even and the other one odd. It is important to note, however, that the series of terms on the RHSs are cut off when the order of the commutator (given by $n$ in $G(j+j'-n)$ in the above formulae\footnote{See  (\ref{fieldinteractions}) which encodes the corresponding interactions among the fields $A_s$.}) exceeds the smallest value of $j$ and $j'$. For most pairs of generators $G(n_q,n_p)$ the RHSs have even fewer terms. For example $G(0)$, which  is a c-number and if added becomes a central element of the HS algebra, does not occur on the RHS of  (\ref{mixedjcommutator}). If one wants to introduce vector fields into this theory it has to be done by extending the spin sum to $A=\Sigma_{s=1}^{\infty}A_s$ but this will not be done  in this paper. In fact, if introduced they would not interact with any of the other fields in $A$
\cite{Pope-Townsend89}. We also see that  the commutator between two operators in the set $G(2)$  gives $G(3)$ from which we conclude  that all values of $j$ will be required to close the  algebra. However, from the above commutator relations it also follows that
the HS algebra can be consistently truncated to consist of only odd spin generators (or even spin fields  in the field theory). 

Once the generators of the HS algebra are defined the next step is to gauge it, that is to introduce one gauge field for each generator. These fields  will thus have spins $s=j+1$. This setup will be described more carefully in later sections. The approximation (or inconsistent truncation) we will adopt in this paper is defined by restricting ourselves to only the first term on the commutator RHS above (or, which is the same, using classical variables and Poisson brackets) {\it and} setting to zero all fields with spin $s=j+1\geq 4$ in the spin 3 analysis in the following two sections and  the same for $s=j+1\geq 5$ in the spin 4 analysis in section 5.

 For these gauge fields, which are divided into spin $s$ sectors with $s=j+1$, we find, for instance, that the spin 4 Cotton equation has interaction terms formed from any two  fields with spins $s=j+1$ and $s'=j'+1$ such that $s+s'-2(n+1)=4$, i.e., $s+s'=(2n+1)+5=6, 8,..$ where $2n+1$ (for $n\geq 0$) is the order of the multi-commutator in the expansion of the star product commutator\footnote{We will here not be precise about the relation between the star product commutator and the Poisson bracket since it will not be needed (it would require the insertion of an $i$ in a number of definitions).}.
These features are of course exactly the ones studied in \cite{Fradkin:1989xt}. The case most relevant in this paper is the one involving spin 2 and spin 3. Then we find that the bilinear interaction terms in the spin 3 Cotton equation contains all terms where the spins of the two fields satisfy $s+s'=(2n+1)+4=5, 7,..$ with $2n+1$ equal to the order of the commutator. The single commutator case $(n=0)$ analysed in this paper therefore 
has only one interaction term containing fields with spin 2 and 3. All equations we will encounter are hence linear in the spin 3 fields. (This is also the case for the spin 2 covariant tensor fields at this stage but after the elimination of the auxiliary spin 3 fields the situation changes drastically as shown in the next section.) In this sense the spin 4 Cotton equation is more interesting since then, in addition to the terms linear in the spin 4 field, there are also terms with no spin 4 fields but with two spin 3 fields. In the truncated even spin case one has to go to the spin 6 Cotton equation for this to happen. 

The  restriction to  terms in the field equations and transformation rules  that arise from single commutators corresponds to expanding the Moyal commutator and keeping only the first term. Note, however, that it is only when using a star product that a consistent HS theory is obtained as demonstrated, e.g.,  in \cite{Nilsson2013} where the Poisson bracket field equations were shown to be incompatible with a Lagrangian formulation. This may be a manifestation of 
problems associated with constructing a non-degenerate bilinear form for the HS Poisson algebra used here.

\subsection{Conformal higher spin Chern-Simons theory}
Three-dimensional conformal gravity can now be obtained from  a gauge theory with gauge group $\mathrm{SO}(2,3)$. In fact the spin 2 gravitational Chern-Simons-like  action is equivalent to the gauge theory Chern-Simons action \cite{Horne-Witten89}
\begin{equation}
  S = \int_M \mathrm{tr}\left( A\wedge \mathrm{d}A + \tfrac{2}{3}A\wedge A\wedge A \right),
  \label{eq:csaction}
\end{equation}
if one identifies the $\mathrm{SO}(2,3)$ gauge (one-form) potential with the  gravitational fields as follows
\beq
\label{gaugefieldspintwo}
A = e_a P^a + \omega_a M^a + b D + f_a K^a.
\eeq
Here $e_a$ is the dreibein, $\omega_a$ the spin connection while $b$ and $f_a$ are (auxiliary) gauge fields for dilatations and special conformal transformations. By giving $q^{\al}$ and $p_{\al}$ dimensions $L^{-\thalf}$ and $L^{\thalf}$, respectively, the one-form $A$ is dimensionless, a fact that continues to hold also  when we let it be  valued in the entire HS algebra below.

In the Chern-Simons formulation of conformal gravity invariance under translation and local Lorentz transformations follow from gauge invariance \cite{Horne-Witten89} 
of the action \eqref{eq:csaction} under gauge the transformations
\begin{equation}
  \mathrm{\delta} A= \mathrm{d} \Lambda + \left[A, \Lambda\right]\; ,
  \label{eq:gaugetransformation}
\end{equation}
 where the gauge parameter $\Lambda$ is an algebra valued zero form.
For conformal gravity the gauge parameter has the form 
\beq
\Lambda = \Lambda^a(2,0) P_a + \Lambda^a(1,1) M_a + \Lambda(1,1) D + \Lambda^a(0,2) K_a,
\eeq
where the $\Lam^a(2,0)$ component generates local translations and the two $(1,1)$ components local Lorentz transformations, $\Lam^a(1,1)$, and dilatations, $\Lam(1,1)$. The last component $\Lam^a(0,2)$ is related to special conformal transformations and contains enough freedom to set the one-form Stückelberg field $b$ to zero, thus leaving the dreibein, spin connection and the Schouten tensor $f_a$ in the theory. Further details of the spin 2 system are given in the review in the next subsection.

To include fields of higher spin we now simply let the one-form gauge field $A$ take values in the entire HS  algebra defined above, i.e.,
\begin{equation}
  A =\sum_{s=2}^\infty A_s= \sum_{s=2}^\infty\sum_{\substack{n_q,n_p\geq 0\\n_q+n_p=2(s-1)}} \sum_{\mr{irreps} (a)} A_s^{(a)}(n_q,n_p)G^{(a)}(n_q,n_p),
\end{equation}
where the sum has been broken up into two parts showing clearly the spin $s$ sector and the irrep content $(a)$ of the terms.
Note that the sum over the irreps $(a)$ (where $(a):=a_1...a_r$ for some integer $r\leq s-1$)  takes care of the fact that given the content of $q$'s and $p$'s the spin may vary as seen from the list of spin 3 generators in (\ref{spinthreegenerators}). The Chern-Simons action in \eqref{eq:csaction} can then be generalized, using the star product,  to  the  higher spin algebra and  the equations of motion stemming from it are 
\begin{equation}
  F = \mathrm{d}A+A\wedge A = 0,
  \label{eq:Fflat}
\end{equation}
i.e., the connection $A$ is flat. This equation can be studied by regarding its irreducible components separately after expanding it in the same way as for $A$ above. In the first step these are given by the different $(q,p)$ components of $F_s$, denoted as  $F_s^{(a)}(n_q,n_p)$ (the index $s$ is usually not written out). For example the component $F^a(2,0)=0$ is the zero torsion condition for the spin 2 connection $\omega_a$ and $F^a(0,2)=0$ is the Cotton equation. 

The HS gauge parameter is generalized in a similar fashion into spin $s$ sectors:
\begin{equation}
  \Lambda = \sum_{s=2}^\infty\Lam_s=\sum_{s=2}^\infty\sum_{\substack{n_q,n_p\geq 0\\n_q+n_p=2(s-1)}} \sum_{\mr{irreps}(a)}\Lambda_s^{(a)}(n_q,n_p)G^{(a)}(n_q,n_p).
\end{equation}
The details of the spin 3 and spin 4 sectors are given in the following sections but as a warm-up we now turn to a review of the spin 2 case.

\subsection{The conformal pure spin 2 system}
The spin 2 part of the higher spin algebra generates the familiar objects of conformal gravity in the first order formalism. For the convenience of the reader we review here some aspects of the spin 2 system relevant for the coming discussions of the spin 3 and 4 systems. Further details of the spin 2 construction can be found in \cite{Horne-Witten89, Nilsson2013}.

Thus we consider the spin 2 sector defined by $A$ given in (\ref{gaugefieldspintwo}). Then taking the spin 2 content of the gauge parameter $\Lambda$ to be\footnote{The notation is designed such that the components of the parameter $\Lam$ and field strength $F$, but not the fields in $A$, need to be accompanied by $(n_q,n_p)$ (written either under or after the quantity in question). This is the case also for spin 3 and 4 treated in later sections.} 
\begin{equation}
  \underset{\text{spin 2}}{\Lambda} = \underset{(2,0)}{\Lambda_a}  P^a + \underset{(1,1)}{\Lambda_a} M^a + \underset{(1,1)}{\Lambda} D + \underset{(0,2)}{\Lambda_a} K^a,
\end{equation}
the gauge transformations  \eqref{eq:gaugetransformation} in this sector read
\[
  \arraycolsep=3.4pt\def\arraystretch{1.5}
  \begin{aligned} 
   \delta\tensor{e}{_{\mu}^{a}} &= \mathrm{D}_{\mu}\underset{(2,0)}{\tensor{\Lambda}{^{a}}} - \tensor{\epsilon}{_{\mu}^{a}^{c}}\underset{(1,1)}{\tensor{\Lambda}{_{c}}} - \tensor{e}{_{\mu}^{a}}\underset{(1,1)}{\tensor{\Lambda}{}} + \tensor{b}{_{\mu}}\underset{(2,0)}{\tensor{\Lambda}{^{a}}},\\
    \delta\tensor{\omega}{_{\mu}^{a}} &= \mathrm{D}_{\mu}\underset{(1,1)}{\tensor{\Lambda}{^{a}}} + 2\tensor{\epsilon}{_{\mu}^{a}^{c}}\underset{(0,2)}{\tensor{\Lambda}{_{c}}} - 2\tensor{\epsilon}{^{a}^{b}^{c}}\tensor{f}{_{\mu}_{b}}\underset{(2,0)}{\tensor{\Lambda}{_{c}}}, \\ 
    \delta\tensor{b}{_{\mu}} &= \mathrm{D}_{\mu}\underset{(1,1)}{\tensor{\Lambda}{}} - 2e_\mu{}^a\underset{(0,2)}{\tensor{\Lambda}{_{a}}} + 2\tensor{f}{_{\mu}^{a}}\underset{(2,0)}{\tensor{\Lambda}{_{a}}}, \\ 
     \delta\tensor{f}{_{\mu}^{a}} &= \mathrm{D}_{\mu}\underset{(0,2)}{\tensor{\Lambda}{^{a}}} - \tensor{b}{_{\mu}}\underset{(0,2)}{\tensor{\Lambda}{^{a}}} + \tensor{\epsilon}{^{a}^{b}^{c}}\tensor{f}{_{\mu}_{b}}\underset{(1,1)}{\tensor{\Lambda}{_{c}}} + \tensor{f}{_{\mu}^{a}}\underset{(1,1)}{\tensor{\Lambda}{}},
  \end{aligned}
\]
where $\mr{D}_\mu$ is the Lorentz covariant derivative containing the spin 2 connection $\omega_{\mu}{}^a$, i.e., 
$De^a=de^a+\ep^a{}_{bc}\om^be^c$.

These transformation rules provide a natural interpretation for the gauge parameters, i.e., $\Lambda^a(2,0)$ parametrizes local translations and $\Lambda_a(1,1)$  local Lorentz transformations while
the remaining parameters $\Lambda(1,1)$ and $\Lambda_a(0,2)$ correspond to scaling transformations and special conformal transformations. Looking at the transformation for $b_\mu$, we note that it is possible to solve for $\Lambda_a(0,2)$ provided the dreibein field $e_{\mu}{}^a$ is declared to be invertible. The subtleties\footnote{For a discussion on this issue, see \cite{Witten:2007kt, Maloney:2007ud}.} associated with this statement will not concern us in this paper. This means that we can use the special conformal transformations to gauge $b_\mu$ to zero and in this gauge we find the familiar formulation of conformal gravity in terms of the dreibein and spin connection as explained below. 
Fields that can be gauged to zero this way will generally be called Stückelberg fields.
Another subtlety related to the  interpretation of this system as three-dimensional conformal gravity concerns  the diffeomorphisms which are manifest symmetries of the Chern-Simons theory. As shown in \cite{Horne-Witten89},  diffeomorphisms can on-shell be identified with a particular linear combination of field dependent gauge transformations. In the next section we will discuss this further  in connection with the spin 3 sector.

Turning to the equations of motion we find that the component equations of $F=0$ for the spin 2 sector of $A$ take the form
\[
  \arraycolsep=3.4pt\def\arraystretch{1.5}
  \begin{aligned}
    \underset{(2,0)}{F^{a}} &= \mathrm{D}_{[\mu}\tensor{e}{_{\nu]}^{a}}  + \tensor{b}{_{[\mu}}\tensor{e}{_{\nu]}^{a}} = 0,\\ 
    \underset{(1,1)}{F^{a}} &=\thalf \tensor{R}{_{\mu\nu}^{a}} + 2\tensor{\epsilon}{_{[\mu}^{ab}}\tensor{f}{_{\nu]}_{b}} = 0, \\ 
    \underset{(1,1)}{F^{}} &= \mathrm{D}_{[\mu}\tensor{b}{_{\nu]}} + 2\tensor{f}{_{[\mu}_{\nu]}} = 0, \\ 
    \underset{(0,2)}{F^{a}} &= \mathrm{D}_{[\mu}\tensor{f}{_{\nu]}^{a}} - \tensor{b}{_{[\mu}}\tensor{f}{_{\nu]}^{a}} = 0,
  \end{aligned}
\]
where we have used the definition of the once dualized Riemann tensor $R_{\mu\nu}{}^a$  in appendix \ref{app:A}.
Thus we see that in the gauge $b_\mu=0$ the first equation reduces to the zero torsion condition, the second says that $f_{\mu}{}^a$ is related to the Schouten tensor
 \beq
 f_{\mu\nu}=\thalf S_{\mu\nu}=\thalf(R_{\mu\nu}-\tfrac{1}{4}g_{\mu\nu}\,R),
 \eeq
the third is a constraint saying that $f_{\mu\nu}$ is symmetric (which is obviously true here) and, finally,  the last equation becomes, after dualization,  the Cotton equation 
\begin{equation}
  C_{\mu\nu}:=\ep_{\mu}{}^{\rho\si}\mathrm{D}_{\rho}f_{\si\nu} = 0,
\end{equation}
solutions of which are conformally flat space-times. This tensor is in the irrep ${\bf 5}$, i.e., it is symmetric and traceless (trivially). It is also  divergence free on both indices. Its transformation properties
are discussed  below.

The content of the component equations of $F=0$ can be summarized as
in the table:
\begin{table}[H]
  \[\arraycolsep=10.4pt\def\arraystretch{1.5}
    \begin{array}{r c c}
      F_{\mu\nu}{}^{(a)}(n_\mathrm{q},n_\mathrm{p})=0 & \mr{so}(1,2) \text{ irreps.} & \text{Solution}\\
      \hline
      F_{\mu\nu}{}^a(2,0)=0 & \pmb{5} \oplus \pmb{3} \oplus \pmb{1} & \omega_\mu{}^a (e_\mu{}^a) \\
      F_{\mu\nu}{}^a(1,1)=0 & \pmb{5} \oplus \pmb{3} \oplus \pmb{1} & f_\mu{}^a (\omega_\mu{}^a) \\
      F_{\mu\nu}{}(1,1)=0 & \con{\pmb{3}} &  \con{\text{Constraint}}\\
      F_{\mu\nu}{}^a(0,2) =0& \pmb{5} \oplus \pmb{3} \oplus \pmb{1} & \text{Cotton eq.}
    \end{array}
  \]
\end{table}

There is a pattern emerging here where some of the equations $F^{(a)}(n_q, n_p)=0$ can be used to solve for \mbox{$A(n_q-1,n_p+1)$}. Using this solution, the last equation, the Cotton equation, is then turned into a third order differential equation for the only independent frame field $e_{\mu}{}^{a}$. As will be clear in the following sections, apart from a few new features, this continues to hold also for higher spins which was first demonstrated at the linear level in \cite{Pope-Townsend89}. In the next two sections we analyze the spin 3 case in detail. 

We  end this section with a brief analysis of the spin 2 symmetries. Imposing the Stückelberg gauge $b_{\mu}=0$ implies (from the equation $\de b_{\mu}=0$) that
\beq
\Lam_{\mu}(0,2)=\thalf \pd_{\mu}\Lam(1,1)+f_{\mu}{}^a\Lam_a(2,0),
\eeq
which then should be inserted into the transformation rules for the other fields given above. We can now check the spin 2 scale invariance of the Cotton tensor:
\beq
\de_{\Lam(1,1)}(\mr{D}_{[\mu}f_{\nu]}{}^a)=\ep^a{}_{bc}(\de\om_{[\mu}{}^b)f_{\nu]}{}^c+\mr{D}_{[\mu}\de f_{\nu]}{}^a=(\mr{D}_{[\mu}f_{\nu]}{}^a)\Lam(1,1)+f_{[\mu\nu]}\pd^a\Lam(1,1),
\eeq
where we used the definitions of the covariant derivative $DV^a=dV^a+\ep^a{}_{bc}\om^bV^c$ and the following relation to the Schouten tensor 
$R_{\mu\nu}{}^{ab}=8\de_{[\mu}^{[a}f_{\nu]}{}^{b]}$ (recall that $f_{\mu\nu}=\thalf S_{\mu\nu}$). Thus we see that $\mr{D}_{[\mu}f_{\nu]}{}^a$ transforms into itself under scalings since $f_{[\mu\nu]}=0$, or  $F(1,1)=0$. The scale invariant quantity is therefore
(see, e.g., \cite{Bunster:2012km}) $
\mr{D}_{[\mu}f_{\nu]}{}^ae_{a\rho}$ since the dreibein scales with weight $-1$. The standard Cotton tensor is defined as $C_{\mu\nu}=\ep_{\mu}{}^{\rho\si}
\mr{D}_{[\rho}f_{\si]}{}^ae_{\nu a}$ and hence has scaling dimension $-2$ (as does the metric).

One may also go to the metric gauge, i.e., impose a Lorentz gauge such that the dreibein is parametrized by a symmetric tensor, the metric $g_{\mu\nu}$. Thus we demand the antisymmetric part of the variation of the dreibein  be zero which implies that the Lorentz parameters are given by
\beq
\Lam_{\mu}(1,1)=-\thalf\ep_{\mu}{}^{\nu\rho}\mr{D}_{\nu}\Lam_{\rho}(2,0).
\eeq
One can then check that the antisymmetric part $f_{[\mu\nu]}$ is not generated by any of the remaining transformations
 (translations and dilatations).
 
 Finally, one may wonder what happens if the spin 3 Cotton tensor is used instead of the spin 2 one.  The spin 3 Cotton tensor is very complicated (see the next two sections) so let us here consider just one single term in it, namely $\mr{D}_{[\mu}f_{\nu]}{}^{ab}$ (with $f_{\nu}{}^{ab}$  defined in (\ref{gaugefieldspinthree}) below) which has spin 2 scaling weight equal to  $+2$. Hence, there seems to exist two spin 2 scaling invariants
 \beq
(\mr{D}_{[\mu}f_{\nu]}{}^{ab})e_{\rho ab},\,\,\,(\mr{D}_{[\mu}f_{\nu]}{}^{ab})e_{\rho a}e_{\si b}.
 \eeq
  Imposing that these two expressions be also spin 3 scaling\footnote{See section \ref{sec:spin3gauge}.} invariant will (most likely) generate the complete spin 3 Lagrangian (the part $[\mu\nu\rho]$) and Cotton equation, respectively. In particular, we know that the spin 3 Cotton equation has more than a thousand terms (if its spin 3 content is expressed in terms of just the spin 3 frame field). This and other aspects of the spin 3 sector will be explained in detail in the next two sections.
\section{The spin 3 sector}
\label{sec:spin3}
The spin 3 content of the connection is given by the expansion
\begin{equation}
\label{gaugefieldspinthree}
  \underset{\text{spin 3}}{A} = e_{a b}P^{a b} + \tilde{e}_{a b} \tilde{P}^{a b} + \tilde{e}_a \tilde{P}^a + \tilde\omega_{a b} \tilde{M}^{a b} + {\tom}_a \tilde{M}^a + \tilde{b} \tilde D + \tilde{f}_{a b} \tilde{K}^{a b} + \tilde{f}_a \tilde{K}^a + f_{a b} K^{a b},
\end{equation}
where we have suppressed the $(n_q,n_p)$ since the notation is unambiguous (compare to $\Lam_3$ below). The one-form fields with two flat indices $ab$ (in the irrep ${\bf 5}$) will be called cascade fields since at the end they will all be determined in a stepwise manner in terms of the frame field $e^{ab}$. The remaining are called auxiliary and are either Stückelberg and/or dependent (if they can be solved for in terms of cascade fields). If a field is both Stückelberg {\it and} dependent the corresponding field equation will become a constraint, i.e., an equation that can be reduced to a relation involving only spin 2 covariant derivatives on the spin 3 frame field and factors of the spin 2 Schouten tensor. Constraints can also arise directly from field equations which can not be used to solve for any field. The full situation 
for the spin 3 system is summarized by tables 1 and 2 below (and for spin 4 in table 4). Note that there is a certain amount of arbitrariness in this procedure, a fact that we will have reason to comment upon later.

Clearly the pattern from spin 2 is repeated but becomes here quite  a bit more involved. However,  no extra difficulties apart from a vastly bigger volume of   computations present themselves. We have therefore developed a Mathematica package to be able to perform these calculations.
The conclusion is that the degrees of freedom that remain after gauge fixing can be solved for in terms of the frame field $e_{a b}$, the spin 3 analogue of the spin 2 dreibein. Again, the last component of the zero field strength condition $F^{ab}(0,4)=0$  becomes the Cotton equation now a fifth order differential equation for the independent degrees of freedom $e_{\mu}{}^{a b}$. This conclusion was reached before for the linearized version of the theory in \cite{Pope-Townsend89} by analyzing the representation content. 
In \cite{Nilsson2013} this was carried out explicitly and there it was also made clear that this could be done  for the non-linear system involving the spin 2 and spin 3 fields. Here we continue in the analysis of this last  paper and solve the full non-linear version of the system.

\subsection{The gauge choice}
\label{sec:spin3gauge}
The spin 3 content of the gauge parameter $\Lambda$ is
\begin{multline}
\label{gaugeparameterspinthree}
  \underset{\text{spin 3}}{\Lambda} = \underset{(4,0)}{\Lambda_{a b}} P^{a b} + \underset{(3,1)}{\Lambda_{a b}} \tilde{P}^{a b} + \underset{(3,1)}{\Lambda_{a}} \tilde{P}^{a} + \underset{(2,2)}{\Lambda_{a b}} \tilde{M}^{a b} + \underset{(2,2)}{\Lambda_{a}} \tilde{M}^{a}
\\ + \underset{(2,2)}{\Lambda} \tilde{D} + \underset{(1,3)}{\Lambda_{a b}} \tilde{K}^{a b} + \underset{(1,3)}{\Lambda_{a}} \tilde{K}^{a} + \underset{(0,4)}{\Lambda_{a b}} {K}^{a b}.
\end{multline}
The resulting gauge transformations \eqref{eq:gaugetransformation} for the spin 3 fields are given in appendix \ref{app:spin3gauge}.
As in the spin 2 case one can set some fields (Stückelberg) to zero by utilizing the symmetries whose transformation rules  contain a shift term\footnote{Terms that only contain a gauge parameter and the spin 2 dreibein.}. In the spin 3 case these are all the symmetries except the  generalized translations ($\Lam^{ab}(4,0)$). Leaving also Lorentz ($\Lam^{ab}(3,1)$) and dilatations ($\Lam^a(3,1)$) symmetries intact one  gauge choice for the remaining symmetries is given in table \ref{tab:gauge}.\footnote{From the spin 3 case summarized in the table \ref{tab:gauge} it seems to be the generators in the (maximal) parabolic subalgebra that is used for this purpose. However, although the spin 3 generators in question
do satisfy $n_p\leq n_q$ and hence are part of the maximal parabolic subalgebra of the entire HS algebra it is only for spin 3 that the Stückelberg symmetries used in the table coincide with these. For spin 2 the Stückelberg ones satisfy 
$n_p < n_q$ while for any spin above 3 the Stückelberg ones extend beyond the maximal parabolic subalgebra and thus do not form a subalgebra. There could, however, exist other reasons for being interested in  restricting  the Stückelberg gauges to the HS parabolic subalgebra.}
\begin{table}[H]
  \[\arraycolsep=3.4pt\def\arraystretch{1.5}
    \begin{array}{r c l l}
    \Lam^{(a)}(n_q,n_p)   & \mr{so}(1,2)\text{ irrep.} & & \text{Interpretation/Gauge} \\
      \hline
      \Lambda^{a b}(4,0)        & \pmb{5} & & \text{Translations ({\small\transl3})} \\
      \Lam^{a b}(3,1)        & \pmb{5} & & \text{Lorentz ({\small\lorentz3})} \\
      \Lam^a(3,1)            & \pmb{3} & & \text{Scale ({\small\scale3})} \\
      \Lam^\irr{a  b} (2, 2) & \pmb{5} & \multirow{3}{*}{$\left.\rule{0cm}{1cm}\right\}$} & \multirow{3}{*}{$\tilde{e}_\mu{}^a = 0$} \\ 
      \Lam^a (2, 2)          & \pmb{3} & &\\
      \Lam (2, 2)            & \pmb{1} & &\\
      \Lam^\irr{a  b} (1, 3) & \pmb{5} & \multirow{2}{*}{$\Bigg\}$} & \multirow{2}{*}{$\tilde{\omega}_\mu{}^a = e_\mu{}^a \hat{\omega}$} \\
      \Lam^{a } (1, 3)       & \pmb{3} & &\\
      \Lambda^\irr{a  b} (0, 4) & \pmb{5} & & \tilde{f}_{\mu}{}^a = \epsilon_{\mu}{}^{a b}\hat{f}_b + e_\mu{}^a\hat{f}\\
    \end{array}
  \]
  \caption{The spin 3 gauge parameters and the Stückelberg gauges used in the text.}
  \label{tab:gauge}
\end{table}
The $\Lam^{ab}(3,1)$ and $\Lam^{a}(3,1)$ parameters are spin 3 generalizations of (spin 2) Lorentz and scale parameters (with one index less). This interpretation stems from the fact they can be used to make the spin 3 frame field $e_\mu{}^{a b}$ both symmetric (Lorentz) and traceless (scale) resulting in a metric like field $\tilde h_{\mu\nu\rho}$\footnote{We save the notation $h_{\mu\nu\rho}$ (without tilde) for the metric with a non-zero trace.}. In a similar fashion it seems natural to view the parameters $\Lam^{ab}(4,0)$ as generalized spin 3 translations. If this means that one can also define spin 3 ``diffeomorphisms''  is not clear since  they do not have a natural action on the coordinates $x^{\mu}$ used to parametrize the spacetime manifold on which the theory is defined. This will be discussed further in section 3.5 below. These spin 3 transformations will be referred to in what follows as \transl3, \lorentz3 and \scale3 as indicated in table \ref{tab:gauge}.

In table 1 we see that all three of the $\Lambda^{(a)}(2,2)$ components are used to set $\tilde{e}_{\mu}{}^a \in \pmb{5}\oplus\pmb{3}\oplus\pmb{1}$ to zero while $\Lambda^{a}(1,3)$ and $\Lambda(1,3)$ are used to gauge away the $\pmb{5}$ and $\pmb{3}$ of $\tilde{\omega}_{\mu}{}^a$. Finally $\Lambda^{ab}(0,4)$ is used to set the $\pmb{5}$ of $\tilde{f}_{\mu}{}^a$ to zero.
Thus we see that the pattern that was present in the spin 2 case where the parameter $\Lambda(n_q, n_p)$ is used to gauge away parts of $A(n_q+1,n_p-1)$ repeats itself here. That this Stückelberg phenomenon is possible stems from the algebra in the following way. 

For the gauge field $A(n_q,n_p)$, its gauge transformation contains terms from $\mathrm{d} \Lambda + \left[ A, \Lambda\right]$ proportional to $G(n_q,n_p)$. In particular a gauge transformation will contain a term proportional to the dreibein $e_a$ if there is a generator $\mathcal{X}$ such that $\left[ P^a, \mathcal{X}\right]$ is proportional to $G(n_q,n_p)$.
Since the commutator lowers both of the $q$- and $p$-degrees by one
\begin{equation}
  \left[G(n_q,n_p), G(m_q,m_p)\right] \propto G(n_q+m_q-1, n_p+m_p-1),
\end{equation}
 it follows that $\left[P^a, \mathcal{X}(q,p)\right] \propto G(q+1, p-1)$. Assuming invertibility of the dreibein and a non-zero commutator this means that $\Lambda(n_q,n_p)$ can be used to gauge away parts of $A(n_q+1, n_p-1)$\footnote{Note that in the star product version of the theory there are no new sources of this phenomenon coming from the multi-commutators.}.

When implementing a partial gauge choice the remaining gauge transformations will be modified. This can be seen for example when the transformation $\Lambda(2,2)$ is used to set $\tilde{e}_a$ to zero, which implies that the parameters $\Lambda(2,2)$ are solved for in terms of $\Lambda(3,1)$ and $\Lambda(4,0)$ (see appending \ref{app:spin3gauge}). This in turn enters into the transformation for $\tilde{e}_{a b}$ which now gets modified. Actually this phenomena never enters calculations when one is working with expressions solely in terms of the fundamental $e_{a b}$ since we are leaving the transformations $\Lambda(3,1)$ and $\Lambda(4,0)$ ungauged. One may nevertheless  wonder about the compatibility of the transformations. It turns out that this works out as it must, that is the transformation of the substituted fields in terms of $e_{a b}$ exactly correspond to the modified gauge transformations of the unsubstituted fields. This again provides a nice way to check the validity of the computer based calculations.
\newpage
\subsection{Zero field strength equations}
The spin 3 content of $A$ (see eq. (\ref{gaugefieldspinthree})) in $F=0$ gives a system that have many similarities with the spin 2 case. The component equations are given in appendix \ref{app:spin3eom}. After implementing the gauge choice discussed above the spin 3 components of $F$ take the following form
\input{eomspin3.tex}
where the fields from the spin two sector, $e_{\mu}{}^a(2,0)$ and $\om_{\mu}{}^a(1,1)$, have been used, respectively, to  convert flat indices to curved ones and construct the covariant derivative $D=d+\om(1,1)$, while the Schouten tensor $f_{\mu}{}^a(0,2)=\thalf(R_{\mu}{}^a-\tfrac{1}{4}e_{\mu}{}^a R)$ appears explicitly. 

The above equations are written in the spin 2 gauge 
$b_{\mu}(1,1)=0$ (see appendix \ref{app:C} for the full  equations prior to implementing any gauge choices). Concerning the spin 3 gauge choice  one could have set to zero  $\tb_{\mu}(2,2)$  instead of the vector part\footnote{This would mean keeping
$\tom_{\mu}{}^a=\ep_{\mu}{}^{ab}\hat\om_b+e_{\mu}{}^a\hat\om$ instead of just the last term.} of  $\tom_{\mu}{}^a(2,2)$ as done here.
The reason we have opted  for the latter possibility is that it simplifies the
equation $F^{ab}(3,1)=0$.\footnote{It is curious to note that in the full star product theory only $\tb_{\mu}(2,2)$ is associated to an operator, namely $\tD(2,2)$, that is non-vanishing when sandwiched between the vacua $|0\rangle_q$ and ${}_p\langle 0|$. Therefore setting $\tb_{\mu}(s-1,s-1)$ to zero for all spins might in general be a more convenient gauge choice.} The corresponding term in $F^{ab}(2,2)=0$ is present (the $\hat f$-term) which will generate some extra terms when solving the above equations in the next subsection (related to the occurrence  of  the operator $\hat{\mathcal O}$ in
(\ref{operatorhatO}) below). This will have consequences in chapter 4 when we compare the frame field formulation to the metric one.

The representation content of this system is summarized in table 2.

\begin{table}[H]
  \[\arraycolsep=10.4pt\def\arraystretch{1.5}
    \begin{array}{r c c c}
      F(n_\mathrm{q},n_\mathrm{p})=0 & \mr{so}(1,2) & \text{Solution} & \text{Constraints}\\
      \hline
      F^\irr{a b}(4,0)=0  & \pmb{7} \oplus \pmb{5} \oplus \pmb{3}  & \tilde{e}^{a b}_\mu \left( e^{a b}_\mu \right) \\
      F^\irr{a b}(3,1)=0  & \pmb{7} \oplus \pmb{5} \oplus \pmb{3}  & \tilde{\omega}^{a b}_\mu \left( \tilde{e}^{a b}_\mu \right) \\
      F^a(3,1)=0  & \con{\pmb{5}} \oplus \pmb{3} \oplus \pmb{1}  &b_{\mu} \text{, }\hat{\omega} & \mathrm{C}_1(\con{\pmb{5}})\\
      F^\irr{a b}(2,2)=0  & \pmb{7} \oplus \pmb{5} \oplus \pmb{3}  &  \tilde{f}^{a b}_\mu \left( \tilde{\omega}^{a b}_\mu \right) \\
      F^a(2,2)=0  & \con{\pmb{5}} \oplus \pmb{3} \oplus \pmb{1}  & \hat{f}_a \text{, }\hat{f} & \mathrm{C}_2(\con{\pmb{5}})\\
      F(2,2)=0  & \con{\pmb{3}} & & \mathrm{C}_3(\con{\pmb{3}})\\
      F^\irr{a b}(1,3)=0 & \pmb{7} \oplus \pmb{5} \oplus \pmb{3} & f^{a b}\left(\tilde{f}^{a b}\right) \\
      F^a(1,3) =0 & \con{\pmb{5}} \oplus \con{\pmb{3}} \oplus \con{\pmb{1}} & & \mathrm{C}_4(\con{\pmb{5}}), \mathrm{C}_5(\con{\pmb{3}}), \mathrm{C}_6(\con{\pmb{1}})\\
      F^\irr{a b}(0,4) =0 & \pmb{7} \oplus \pmb{5} \oplus \pmb{3} & \text{Cotton eq.}

    \end{array}
  \]
 \caption{Representation content of the spin 3 equations of motion and the solution cascade. The $F=0$ in the underlined representations are not used in solving for any of the fields and hence become constraints $C_i$ ($i=1,...,6$). The Stückelberg fields have been set to zero here.}
  \label{tab:eom}
\end{table}

From table 2 we see that there is a set of solutions arising from solving all but the last of the zero field strength equations in the repr ${\bf 5}$, i.e., $F^{ab}(n_q,n_p)=0,\,n_q>0$, which we will  refer to
as the {\it cascade}:
\beq
f_{\mu}{}^{ab}\rightarrow \tf_{\mu}{}^{ab}\rightarrow \tom_{\mu}{}^{ab}\rightarrow \te_{\mu}{}^{ab}\rightarrow e_{\mu}{}^{ab}.
\eeq
Here the arrows indicate that the solution gives the field to the left of the arrow as a function, containing one derivative, of the field to the right of the arrow plus some non-linear terms involving  other spin 3 fields further down the cascade multiplied by the spin 2 Schouten tensor(s). The explicit solution is presented in full detail in the next subsection. We also see from table 2 that the rest of the fields in $A_3$ not set to zero by the gauge choice can also solved for in terms of cascade fields (not indicated in table 2). The remaining components of $F=0$ can not be used to solve for any fields and will thus become a subset of the {\it constraints} as discussed further below. The remaining constraints arise from the components of $F=0$ that allows one to solve for a Stückelberg field which is then gauged to zero. The structure of the whole set of equations in $F=0$ described here for spin 3 is generic and arise for all values of $s$. It is basically just a result of counting irreps as was explained (in the linearized setting) by Pope and Townsend in \cite{Pope-Townsend89}.

In the next subsection we present the full solution to the spin 3 component equations.
With the solution at hand we can then turn to the two remaining issues: verification of the  constraints in subsection 3.4 and the structure of the spin 3 Cotton equation in section 4.
\subsection{The solution}
\label{sec:solution}

It is now a trivial, but somewhat tedious, matter to solve the non-linear spin  2 covariant spin 3 component equations (except, of course, the Cotton equation $F_{\mu\nu}{}^{ab}(0,4)=0$)
 of $F_3=0$ given in the previous subsection. The procedure for doing this should be clear from table 2. 
Just as in the spin 2 case there is a solution cascade where one uses $F(n_q,n_p)=0$ to solve for parts of $A(n_q-1,n_p+1)$ giving it as a sum of  terms with a spin 2 covariant derivative acting on parts of $A(n_q,n_p)$ plus some non-linear terms involving both the covariant tensor $f_{\mu}{}^a(0,2)$ from the spin 2 sector and spin 3 fields further down in the cascade. This pattern arises for exactly the same reason as  in the case of the gauge transformations.

The solution reads
\begingroup
\allowdisplaybreaks
\input{spin3solutions.tex}
\endgroup
By inspecting these equations we see that all the fields appearing in the spin 3 sector can be 
expressed in terms of the basic field in this sector, the frame field $e_{\mu}{}^{ab}$. Returning to the spin 3 Cotton equation $F^{ab}(0,4)=0$ given in the previous subsection we can see why  substituting  the solution into it will result in  a tremendously 
complicated equation. We will return to this equation in section 4.
\subsection{Constraints}

From the representation content (see, e.g., table \ref{tab:eom}) of the equations of motion $F_3=0$ it is clear, as explained above, that some of the equations become constraints on the system. These constraints are collected in table \ref{tab:constraints}.
\begin{table}[H]
  \[\arraycolsep=10.4pt\def\arraystretch{1.5}
    \begin{array}{c l}
      \mathrm{C}_1 & F^a(3,1)\repcontent{5}=0 \\
      \mathrm{C}_2 & F^a(2,2)\repcontent{5}=0 \\
      \mathrm{C}_3 & F(2,2)\repcontent{3}=0 \\
      \mathrm{C}_4, \mathrm{C}_5, \mathrm{C}_6 & F^a(1,3)\repcontent{5,3,1}=0
    \end{array}
  \]
  \caption{Constraints from the remaining equations in table \ref{tab:eom}.}
  \label{tab:constraints}
\end{table}

One expects that given the cascading solution of $f^{a b}$ and the other dependent fields in terms of $e^{a b}$ these constraints are all identities. This is indeed the case as will now be shown. The further down one looks in the list of constraints the more of the solution cascade needs to be used and hence more work needs to be done to verify the equation. In the linear case, some of the constraints were checked in \cite{Nilsson2013}. As a non-linear  example we look a bit closer at $\mathrm{C}_5$.

This constraint consists of the $\pmb{3}$ of $F_{\mu\nu}{}^a(1,3)$ (here written with its form indices). Contracting it with a dreibein picks out the $\pmb{3}$ which reads
\begin{equation}
  \left(\mathrm{C}5\right)_\nu = e^{\mu}{}_a F_{\mu \nu}{}^a(1,3) = 3f_{a \nu}{}^{a} + 2\tilde{b}_{[\nu} f_{a]}{}^{a} + f_{\nu [a}\tilde{\omega}_{b]}{}^{a b}-\tfrac{1}{2} \epsilon_\nu{}^{a b}\mathrm{D}_a \hat{f}_b-\mathrm{D}_\nu \hat{f}.
\end{equation}

As explained above, substituting the spin 3 solution cascade $f_\mu{}^{a b}\rightarrow\tilde{f}_\mu{}^{a b}\rightarrow\tilde{\omega}_\mu{}^{a b}\rightarrow\tilde{e}_\mu{}^{a b}\rightarrow e_\mu{}^{a b}$ results in a rather big expression but now only in terms of the spin 3 frame field $e_\mu{}^{a b}$ and the spin 2 Schouten tensor $f_\mu{}^a$. 

Since every step in the substitution involves terms which introduce one more derivative and terms that don't, there exists terms with zero, two or four derivatives. At this point the expression consists of different contractions from the set
\begin{equation}
  \left\{ 
    f_\mu{}^{a}f_\nu{}^{b}e_\rho{}^{c d}, 
    \mathrm{D}_\mu\mathrm{D}_\nu f_\rho{}^{a} e_\sigma{}^{b c},
    f_\mu{}^{a} \mathrm{D}_\nu\mathrm{D}_\rho e_\sigma{}^{b c},
    \mathrm{D}_\mu f_\nu{}^a \mathrm{D}_\rho e_\sigma{}^{b c}, 
    \mathrm{D}_\mu\mathrm{D}_\nu \mathrm{D}_\rho\mathrm{D}_\sigma e_\tau{}^{b c} \right\}.
\end{equation}
Let us start by considering the terms of highest order in derivatives in this list. It turns out that the terms containing symmetrized or contracted derivatives drop out. This feature was seen for the first four constraints already in \cite{Nilsson2013} where they
were shown to be identities at the linear level. At the non-linear level, working with covariant derivatives, this fact can be established by choosing a fixed lexicographic ordering of the derivatives. Then collecting all terms results in a complete cancellation of terms with four derivatives. Thus the $C_5$ constraint can  be rewritten as lower order terms together with spin 2 curvature tensors which in turn can be related to the Schouten tensor by $ f_\mu{}^{a} = \tfrac{1}{2}\left(R_\mu{}^a-\tfrac{1}{4}e_\mu{}^a R\right)$. Repeating the lexicographic step for the terms with two derivatives acting on the same field/tensor these can also seen to vanish.  So  in the end all terms are different contractions of just two kinds of structures, namely
\begin{equation}
  \left\{ f_\mu{}^{a}f_\nu{}^{b}e_\rho{}^{c d}, 
  \mathrm{D}_\mu f_\nu{}^a \mathrm{D}_\rho e_\sigma{}^{b c}\right\}.
  \label{eq:c5terms}
\end{equation}
Finally the spin 2 Cotton equation  together with the Bianchi identity $\mr{D}_a f_\mu{}^a = \mr{D}_\mu f_a{}^a$ can be used to show that the remaining terms indeed sum up to zero. 

These calculations quickly becomes unwieldy and so have been mostly carried out with the help of a computer algebra system.\footnote{There are many excellent tensor algebra systems today however the somewhat specialized nature of this problem made a custom solution in Mathematica together with the index canonicalization code xPerm \cite{xperm} the approach of choice.} In particular in the last step where the many possible contractions ensues the use of Schouten type identities (i.e., ``cycling of indices'') it has been very helpful to verify the status of the expression by explicit index calculations. 

In fact, since the constraints are expected to be identities they provide an excellent testing ground when developing the Mathematica techniques for dealing with this system of equations. The main purpose for doing this is of course to apply them to the spin 3 Cotton equation which at the non-linear level contains more than a thousand terms.

\subsection{Spin 3 ``metric'' and  ``diffeomorphisms''?}

The  formalism used  in this paper to construct a conformal HS theory in $2+1$ dimensions starts from a HS algebra based on $\mr{so}(3,2)$, which is then gauged by introducing a one-form gauge potential $A$ valued in the HS algebra. Using this gauge field one can then as usual write down the Chern-Simons  Lagrangian. A HS theory constructed in this way has of course  both the gauge symmetry of the HS algebra and the diffeomorphism invariance of the (topological) Chern-Simons gauge theory. For the pure spin 2 system (based on the ordinary $\mr{so}(3,2)$ algebra) it is  standard  to define  diffeomorphisms in terms of the  $\mr{so}(3,2)$ gauge symmetries in order to avoid having two infinitesimal symmetries that act in an identical manner on the fields. This is done as follows \cite{Horne-Witten89}
\beq
\de x^{\mu}=\xi^{\mu}(x):\,\,\,\,\de_{\mr{diff}_2(\xi^{\mu})}=\de_{\hat\Lam^a(2,0)}+\de_{\hat\Lam^a(1,1)}+\de_{\hat\Lam(1,1)}+\de_{\hat\Lam^a(0,2)}\,\,\,\,\,\mr{mod}(F=0),
\eeq
where $\de_{\mr{diff}_2(\xi^{\mu})}$ denotes an ordinary infinitesimal coordinate transformation (in $2+1$ dimensions) and the $\Lam$ parameters are given by the field dependent expressions
\beq
\hat\Lam^a(2,0)=\xi^{\mu}e_{\mu}{}^a,\,\,\,\,\,\hat\Lam^a(1,1)=\xi^{\mu}\om_{\mu}{}^a,\,\,\,\,\,
\hat\Lam(1,1)=\xi^{\mu}b_{\mu}{},\,\,\,\,\,\hat\Lam^a(0,2)=\xi^{\mu}f_{\mu}{}^a.
\eeq
That is, on-shell linear coordinate transformations can  be identified  as a  field dependent combination of translations ($\Lam^a(2,0)$), Lorentz ($\Lam^a(1,1)$), dilatation ($\Lam(1,1)$) and special conformal  transformations ($\Lam^a(0,2)$).
In verifying that this combination of $\mr{so}(3,2)$ symmetries gives exactly linearized diffeomorphisms  one has to impose the whole set of $\mr{so}(3,2)$ field strengths components equal to zero.
This kind of procedure also works for other Chern-Simons gravity (i.e., spin 2) systems which are not conformal, e.g., when the cosmological constant is either positive, zero or negative. These conclusions are not affected by the gauge choice $b_{\mu}=0$ imposed in the spin 2 sector. One should, however, be aware of the fact that non-linearly
this identification of diffeomorphisms as a combination of gauge symmetries seems to meet with considerable problems\footnote{ One of the authors (BEWN) thanks M. Duff for discussions concerning this issue.}.

The natural question to ask is then if the same kind of combination of spin 3 transformations can be given a  concrete geometric interpretation similarly to that in the spin 2 case. I.e., can
\beq
\sum_{m,n}^{m+n=4}\de_{\hat\Lam(m,n)}\,,
\eeq
with spin 3 field dependent parameters $\hat\Lam^{ab}(4,0)=\xi^{\mu}e_{\mu}{}^{ab}(4,0)$, etc, be related to any kind of coordinate transformation? The answer is that the result we reviewed above for spin 2 must generalize to the whole HA algebra. Thus by imposing the HS equations $F=0$ the identification between diffeomorphisms and translations in the HS gauge algebra works also for the HS fields with spin $s>2$. However, if one is asking for the effect of these HS spin gauge symmetries on the dreibein the answer seems to be ``no'' but issues that may be connected to this and indicate a different answer have been discussed in the literature, see, e.g., \cite{Ammon:2011nk}, where spin 3 transformations do affect the metric and are used to eliminate physical singularities in the geometry (as defined by the spin 2 metric).

In fact,  the complete effect of all higher spin translation transformations on the spin 2 dreibein can be determined using the star product and is given by the  terms containing $\Lam^{a_1..a_{s-1}}_s(2s,0)$ in the following expression
\beq
\de_{HS} e_{\mu}{}^a=\mr{D}_{\mu}\Lam^a(2,0)+\sum_{s= 3}^{\infty}\sum_{n_q,n_p\geq 0}\sum_{n'_q,n'_p\geq 0}C^{a,(b),(c)}_s(n_q,n_p;n'_q,n'_p)\,A_s^{(b)}
(n_q,n_p)
\Lam^{(c)}_s(n'_q,n'_p),
\eeq
where the coefficients $C^{a,(b),(c)}_s(n_q,n_p;n'_q,n'_p)\neq 0$ when
$n_q+n_p=n'_q+n'_p$ 
(i.e., $s=s'$ which has already been implemented in the sums)
and $(n_q+n'_q,n_p+n'_p)=(2,0)+(2m+1,2m+1)$ where $2m+1$ 
 $(m\geq0)$ is the order of the multi-commutator used to get the answer. Note, however, that the HS translation parameters may also arise from some of the other parameters after gauge fixing\footnote{This is only true in the non-linear theory as will be clear in the next section.}. The total HS variation of the spin 2 dreibein is thus  very complicated  and it is not clear what it is trying to tell us.

In this context one should note that in Chern-Simons theories giving  a HS theory with spins $2,...,N$ in $\mr{AdS}_3$ it is possible to define a HS invariant metric by using the trace in the HS algebra $\mr{sl}(N,\mathbb{R
})\oplus \mr{sl}(N,\mathbb{R})$. Thus, by defining the invariant metric by $\hat g_{\mu\nu}=\mr{Tr}(e_{\mu}e_{\nu})$ \cite{Campoleoni:2010zq}, where $e_{\mu}$ is the frame field constructed by summing over all ``translation'' operator valued fields in the $\mr{sl}(N,\mathbb{R})\oplus \mr{sl}(N,\mathbb{R})$ theory,
one can design HS frame fields that generate drastic changes in the geometry depending on whether it is described by $g_{\mu\nu}=e_{\mu}{}^ae_{\nu}{}^b\eta_{ab}$ or $\hat g_{\mu\nu}=\mr{Tr}(e_{\mu}e_{\nu})$. For examples, see \cite{Ammon:2011nk}. The issue of defining the metric and its spin 3 analogue from the $\mr{sl}(3,\mathbb{R})\oplus \mr{sl}(3,\mathbb{R})$ frame fields is afflicted by some really hard problems as can be seen in recent work \cite{Ammon:2011nk, Fredenhagen:2014oua}. 

The problem of how to define a 
non-degenerate HS frame field in general in $AdS_3$ was addressed in \cite{Fujisawa:2012dk} and a criterion defined which was later used in \cite{Lei:2015ika} to discriminate between various background solutions. Interestingly enough, among the possible Lifshitz and Schroedinger solutions with dynamical exponent $z$ checked in \cite{Lei:2015ika} only the latter
with  $z=2$ was found to pass the test. This  coincides with the
 result of \cite{Nilsson:2013fya} where  solutions of the topologically gauged $CFT_3$ spin 2 Cotton and Klein-Gordon equations are discussed.

In our conformal case there is no dimensionful parameter so the above definition does not exist (in the unbroken theory)\footnote{Recall that the dimension of the spin $s$ frame field $e_{\mu}{}^{a_1...a_{s-1}}$ is $L^{s-2}$.}. Instead one may consider expressions like 
\beq
\hat g_{\mu\nu}=\mr{Tr}(e_{(\mu}f_{\nu)}),
\eeq
where one sum over frame fields has been replaced (for each spin) by a dual field with maximal number of $p$  instead of $q$ operators\footnote{In the spin 2 sector this dual field is the Schouten tensor field $f_{\mu}=f_{\mu}{}^a(0,2)K_a(0,2)$.}. In the star product theory the trace is defined by $\mr{Tr}(\phi(q,p))=\phi(0,0)$ \cite{Fradkin:1989xt, Vasiliev:1989re}, which gives the proper reason why the usual definition does not make sense: $\mr{Tr}(e_{\mu}e_{\nu})=0$. The definition $\hat g_{\mu\nu}=\mr{Tr}(e_{(\mu}f_{\nu)})$ gives a ``metric'' with dimension $L^{-2}$ and thus seems not to be correct. However, for the conformal theory in an Einstein background this definition could nevertheless produce a sensible metric since  one could then replace the spin 2 Schouten tensor $f_{\mu}{}^a$ by $\tfrac{1}{4}\Lambda e_{\mu}{}^a$ and drop the cosmological constant dependent factor. This may then at the end give  a metric definition similar  to the one used 
in the $\mr{sl}(N,\mathbb{R})\oplus \mr{sl}(N,\mathbb{R})$ type HS theories mentioned above although it is not clear how to deal with the terms with spin larger than two.

\section{The spin 3 Cotton equation}
In this section we analyze the spin 2 covariant spin 3 Cotton equation $F^{ab}(0,4)=0$ from various perspectives.
As should be clear from the previous discussions in this paper, writing out  this equation in full detail in terms of the basic frame field $e_{\mu}{}^{ab}(4,0)$ at the non-linear (in spin 2 fields) level is doable (as a computer output \cite{LinanderMath}) but hardly very useful for the presentation or calculations by hand. Therefore, this will not be done here but we hope to remedy this in a future publication. At the moment, the strategy will instead be to provide a simplified but to some extent implicit presentation of the spin 3 Cotton equation. This way we may analyze it in different ways as we now explain.

In the first subsection below we discuss
the linearized version aiming at connecting
our results to previous ones in the literature. Since most of these are in the ``metric'' formulation we will  be forced to define exactly the relation between the spin 3 frame field and the ``metric''. 

The second subsection is then devoted to a discussion of the symmetry properties of the various fields that appear in expansion of $A_3$ after they have been expressed in terms of the frame field $e_{\mu}{}^{ab}$. In particular we compare the linearized case to the fully non-linear situation since this has implications for the transition to the metric formulation. 

Finally, in the last subsection we turn to the full non-linear spin 3 Cotton equation and present
the exact equation in a form that can be written out in a compact way. This is achieved by splitting it up into two parts. The first part is obtained by repeating the cascading procedure of the first subsection but now using 
 covariant instead of partial derivatives in the various operators defined there.  The second part contains then
the remaining terms which all involve the spin 2 Schouten (or Ricci) tensor.
Presenting this latter set of terms would be easier since the cascading has a smaller number of steps than the first part of the Cotton equation. However, we will not present any of these parts  in detail but hope to return to this issue in the future. Combining the results obtained in this section it will be obvious how to extract the fully non-linear spin 3 Cotton equation in the metric formulation. This (together with the frame field result \cite{LinanderMath}) is one of the main results of this paper and is presented here as an output from our computer algebra aided computations \cite{LinanderMath2}.

\subsection{Linearized theory and connection to the ``metric'' formulation}

One of the goals of this subsection will be to verify  that our expressions possess the expected symmetry properties connected to the remaining unfixed gauge parameters. These symmetries, together with the spin 3 Ricci, Schouten and Cotton tensors, were  discussed in detail some time ago
by Damour and Deser \cite{Damour:1987vm}  (for some more recent discussions, see  Bergshoeff et al. \cite{Bergshoeff:2009tb, Bergshoeff:2011pm} and Henneaux et al. \cite{Henneaux:2015cda}) but then in the linearized ``metric'' formulation\footnote{Linearized unfolded equations for HS fields in $AdS_3$  were classified in  \cite{Boulanger:2014vya} where also the conformal systems were discussed.}. 
  We therefore end this subsection by giving the precise connection between the spin 3 frame field that appear  here and the ``metric'' of these previous references and in the process we show that the ``metric'' formulation of the linearized tensors above as well as the 
spin 3 Cotton equation derived  in \cite{Damour:1987vm} 
are reproduced by our formalism.

The set of cascade equations we are dealing with here is
{
\setlength{\jot}{1.5ex}
\begin{align}
\te_{\mu}{}^{ab}|_{\mr{lin}}&=\ep_{\mu}{}^{\nu\rho}\pd_{\nu}e_{\rho}{}^{ab}-2(\ep^{\nu\rho(a}\pd_{\nu}e_{\rho\mu}{}^{b)}-\tfrac{1}{3}\eta^{ab}\ep^{\nu\rho c}\pd_{\nu}e_{\rho \mu c}),\\
\tom_{\mu}{}^{ab}|_{\mr{lin}}&=-\thalf(\ep_{\mu}{}^{\nu\rho}\pd_{\nu}\te_{\rho}{}^{ab}|_{\mr{lin}}-2(\ep^{\nu\rho(a}\pd_{\nu}\te_{\rho\mu}{}^{b)}|_{\mr{lin}}-\tfrac{1}{3}\eta^{ab}\ep^{\nu\rho c}\pd_{\nu}\te_{\rho \mu c}|_{\mr{lin}})),\\
\tf_{\mu}{}^{ab}|_{\mr{lin}}&=
\tfrac{1}{3}(\ep_{\mu}{}^{\nu\rho}\pd_{\nu}\tom_{\rho}{}^{ab}|_{\mr{lin}}-2(\ep^{\nu\rho(a}\pd_{\nu}\tom_{\rho\mu}{}^{b)}|_{\mr{lin}}-\tfrac{1}{3}\eta^{ab}\ep^{\nu\rho c}\pd_{\nu}\tom_{\rho \mu c}|_{\mr{lin}}))\\
&\phantom{=} -\tfrac{1}{48}(\ep^{\nu\rho c}e_{\mu}{}^{(a}\pd_{\nu}\tom_{\rho c}{}^{b)}|_{\mr{lin}}-\tfrac{1}{3}\eta^{ab}\ep^{\nu\rho c}\pd_{\nu}\tom_{\rho c \mu }|_{\mr{lin}})),\\
f_{\mu}{}^{ab}|_{\mr{lin}}&=-\tfrac{1}{4}(\ep_{\mu}{}^{\nu\rho}\pd_{\nu}\tf_{\rho}{}^{ab}|_{\mr{lin}}-2(\ep^{\nu\rho(a}\pd_{\nu}\tf_{\rho\mu}{}^{b)}|_{\mr{lin}}-\tfrac{1}{3}\eta^{ab}\ep^{\nu\rho c}\pd_{\nu}\tf_{\rho \mu c}|_{\mr{lin}})),
\end{align}
}
which is the  linearized version of the corresponding equations in (\ref{tesolution}) to (\ref{fsolution}) in the sense that the spin 2 metric has been set equal to the Minkowski one. The expressions we obtain this way are the proper ones for comparison to the results of \cite{Damour:1987vm, Bergshoeff:2009tb} and \cite{Henneaux:2015cda}.
The linearized Cotton equation  $F^{ab}(0,4)|_{\mr{lin}}=0$  after dualization in the frame field formulation reads
\beq
C_{\mu}{}^{ab}|_{\mr{lin}}:=\ep_{\mu}{}^{\rho\si}\pd_{\rho}f_{\si}{}^{ab}|_{\mr{lin}}=0,
\eeq
where $f_{\mu}{}^{ab}|_{\mr{lin}}$ should be expressed in terms of the frame field 
$e_{\mu}{}^{ab}$ by means of the above cascade equations.

To express this equation in an as simple form as possible  we now introduce three different  operators constructed from one epsilon tensor and one partial derivative (which will become spin 2 covariant in the next subsection) 
mapping from the space of frame tensors (that is tensors like $e_{\mu}{}^{ab}$) to itself. These operators, which do not commute,  are denoted $\mcO, \mchO, \sep$ and are defined by the cascade equations above as  follows:
\beq
\label{operatorO}
\mcO:\,\,\,e_{\mu}{}^{ab} \rightarrow \te_{\mu}{}^{ab}|_{\mr{lin}}=({\mathcal O}e)_{\mu}{}^{ab},
\eeq
\beq
\label{operatorhatO}
\mchO:\,\,\,\tom_{\mu}{}^{ab}|_{\mr{lin}} \rightarrow \tf_{\mu}{}^{ab}|_{\mr{lin}}=\tfrac{1}{3}({\mathcal O}\tom|_{\mr{lin}})_{\mu}{}^{ab}+\tfrac{1}{3}(\mathcal {\hat O}\tom|_{\mr{lin}})_{\mu}{}^{ab},
\eeq
and finally
\beq
\label{epslash}
\sep: \,\,\,\,(\sep)_{\mu}{}^{\nu}:=\ep_{\mu}{}^{\rho\nu}\pd_{\rho}\,,
\eeq
acting as follows on the frame field tensors
\beq
(\sep e)_{\mu}{}^{ab}=(\sep)_{\mu}{}^{\nu}e_{\nu}{}^{ab}=\ep_{\mu}{}^{\rho\nu}\pd_{\rho}e_{\nu}{}^{ab}.
\eeq
Note that the last operator appears both in the linearized Cotton equation acting on $f_{\mu}{}^{ab}$ and in the first term of the operator 
${\mathcal O}$.

With the notation introduced above we can now easily use the cascade equations to express the field $f_{\mu}{}^{ab}|_{\mr{lin}}$ in terms of the basic frame field $e_{\mu}{}^{ab}$. The linearized Cotton equation then takes the simple form
\beq
\label{framecotton}
C_{\mu}{}^{ab}(e)|_{\mr{lin}}:=(\sep f(e)|_{\mr{lin}})_{\mu}{}^{ab}=\tfrac{1}{4!}\sep(\mcO^4e+\mcO\mchO\mcO^2e)_{\mu}{}^{ab}=0.
\eeq

Later when we turn the spin 3 Cotton equation into an equation for the spin 3 metric $h_{\mu\nu\rho}$ we need to address the following questions: is the spin 3 Cotton tensor defined as
\beq
C_{\mu\nu\rho}(h)|_{\mr{lin}}:=(\sep f(h)|_{\mr{lin}})_{\mu\nu\rho},
\eeq
i) symmetric in all three indices,
ii) traceless on any pair of indices,
iii) divergence free on all indices as expected? 
Note that if the first question can be answered in the affirmative the other two properties follow since this tensor is by definition symmetric and traceless in the last two indices and divergence free on the first one due to its relation to the frame field formulation. The reason we raise these questions is that in previous metric formulations, see \cite{Damour:1987vm, Bergshoeff:2009tb}, these properties are known to be automatically true \cite{Henneaux:2015cda}. We also need to answer these questions for the full non-linear Cotton tensor. This will be done later in this section.

At this point, however,  we are still in the frame field formulation and since all three operators $\mcO, \mchO, \sep$ defined above map the space of tensors which are symmetric and traceless in  the last two indices 
into itself also the Cotton tensor will belong to this space of tensors. This is in fact true also with the spin 2 covariantized  operators used in the next subsection as well as for the full non-linear theory. This follows  directly from the cascade equations. We note that the Cotton tensor $C_{\mu}{}^{ab}(e)$ is also trivially divergence free on the first index a fact that continues to hold in the non-linear spin 2 covariant case due to the gauge theory Bianchi identity\footnote{One should note, however, that the Bianchi identity $DF^{ab}(0,4)=0$ after the elimination of the auxiliary fields is only satisfied modulo the solution of the spin 2  cascade equations.}. We will continue this discussion after we have converted the Cotton equation into its metric form.

In order to get a feeling for what kind of expressions that will occur  we give here the linearized cascade field $\tom_{\mu}{}^{ab}|_{lin}=-\thalf({\mathcal O}^2e)_{\mu}{}^{ab}$ explicitly:
\begin{multline}
\label{tomlinear}
\tom_{\mu}{}^{ab}|_{\mr{lin}}=-\thalf(-\Box e_{\mu}{}^{ab}+\pd_{\mu}\pd^{\nu}e_{\nu}{}^{ab}-\tfrac{4}{3}e_{\mu}{}^{(a}\pd_{\nu}\pd_{\rho}
e^{|\nu\rho|b)}+\tfrac{4}{3}e_{\mu}{}^{(a}\Box e_{\nu}{}^{b)\nu}+2\pd^{(a}\pd_{\nu}e_{\mu}{}^{b)\nu}
+2\Box e^{(ab)}{}_{\mu}\\
-2\pd^{\nu}\pd^{(a}e_{\nu}{}^{b)}{}_{\mu}-
2\pd^{\nu}\pd^{(a}e^{b)}{}_{\nu}{}_{\mu}+2\pd^a\pd^be_{\nu\mu}{}^{\nu}
-2\pd_{\mu}\pd^{\nu}e^{(ab)}{}_{\nu}+\tfrac{4}{3}e_{\mu}{}^{(a}\pd_{\nu}\pd_{\rho}
e^{b)\nu\rho}-\tfrac{4}{3}e_{\mu}{}^{(a}\pd^{b)}\pd_{\nu}e_{\rho}{}^{\nu\rho}\\
+\tfrac{16}{9}\eta^{ab}(\pd_{\nu}\pd_{\rho}e^{\nu\rho}{}_{\mu}-\Box e_{\nu\mu}{}^{\nu})
+\tfrac{10}{9}\eta^{ab}(-\pd_{\nu}\pd_{\rho}e_{\mu}{}^{\nu\rho}+\pd_{\mu}\pd_{\nu} e_{\rho}{}^{\nu\rho})).
\end{multline}
It is trivial to check that the expression on the right hand side is traceless in $ab$ as it must. One may also verify that it is \transl3 invariant under\footnote{This is trivially true also for $\te_{\mu}{}^{ab}|_{\mr{lin}}$.} 
\beq
\de e_{\mu}{}^{ab}|_{\mr{lin}}=\pd_{\mu}\Lam^{ab}(4,0),
\eeq
but, as expected, not \scale3 invariant. That $\de\tom_{\mu}{}^{ab}|_{\mr{lin}}$ depends on $\Lam^a(3,1)$ can be seen from the linearized version of the transformation rules in appendix \ref{app:C2} (obtained by setting $f_{\mu}{}^a(0,2)=0$). By inspecting these equations (which  involve only the transformation rules for fields in irrep $a$) we see that all the parameters that are solved for at this stage become, schematically,  equal to $\pd\Lam^a(n_q,n_p)$ for some $n_q,n_p$. Thus the sequence 
$\Lam^a(1,3)\rightarrow \pd\Lam^a(2,2)\rightarrow \pd\pd\Lam^a(3,1)$ tells us that there is always a potential dependence of the spin 3 scale parameter as soon as any of these parameters appear in the transformation rules in appendix \ref{app:C2}. Whether or not the corresponding symmetries are actually present must, however, be checked explicitly in each case.

The first linearized cascade field that is \scale3 invariant is, in fact, the Cotton 
tensor\footnote{Strictly speaking the Cotton tensor is not a cascade field since it is not one of the fields in $A_3$.} which we have verified using our Mathematica package, i.e., $f_{\mu}{}^{ab}(0,4)|_{\mr{lin}}$ expressed in terms of $e_{\mu}{}^{ab}(4,0)$ is not invariant. This is in accord with a theorem proved for the linear metric theory in \cite{Damour:1987vm}. For some more recent work discussing this point, see \cite{Bergshoeff:2009tb, Henneaux:2015cda}. In the frame field formulation 
this is true also for the non-linear theory analyzed in this paper since the Cotton equation  is just a component of the zero field strength equation (but we have also verified it explicitly using Mathematica). Note, however, that this conclusion,
as will be clear from the discussion in the last subsection, also holds for the metric formulation which follows from our derivation of it from the frame field theory.

Turning to the next cascade field $\tf_{\mu}{}^{ab}(1,3)|_{\mr{lin}}$ we saw  in subsection \ref{sec:solution} that there is an extra derivative term in its relation to $\tom_{\mu}{}^{ab}(2,2)|_{\mr{lin}}$ which is not there in the other cascade relations. As we will now argue this  is related to the form of the Schouten tensor defined  in \cite{Bergshoeff:2009tb} (see also \cite{Henneaux:2015cda}).
In the notation of \cite{Bergshoeff:2009tb} the linearized Schouten tensor, written  in terms of their spin 3 ``metric'', or Fronsdal field, $h_{\mu\nu\rho}$, is
\beq
\label{schoutenthree}
S_{\mu\nu\rho}(h)|_{\mr{lin}}=G_{\mu\nu\rho}(h)|_{\mr{lin}}-\tfrac{3}{4}\eta_{(\mu\nu}G_{\rho)}(h)|_{\mr{lin}},
\eeq
where  the spin 3 ``Einstein'' tensor\footnote{Note that although there are three $\sep$ operators they do not act as in the cascade field $F_{\mu}{}^{ab}$ where they come from $\mathcal O$  and thus appear as $\sep^3$ acting only on the curved index of $e_{\mu}{}^{ab}$!}
\beq
G_{\mu\nu\rho}|_{\mr{lin}}:=-\tfrac{1}{6}(\sep)_{\mu}{}^{a}(\sep)_{\nu}{}^{b}(\sep)_{\rho}{}^{c}h_{abc},
\eeq
now written in terms of our definition of $(\sep)_{\mu}{}^{a}$ in (\ref{epslash}). Here we have also defined the  trace $\eta^{\nu\rho}G_{\mu\nu\rho}|_{\mr{lin}}:=G_{\mu}|_{\mr{lin}}$ which gives that $\eta^{\nu\rho}S_{\mu\nu\rho}|_{\mr{lin}}=-\tfrac{1}{4}G_{\mu}|_{\mr{lin}}$.

From this definition of the Einstein tensor we see that it is automatically symmetric and  divergence free as well as \transl3 invariant since $(\sep)_{\mu}{}^{a}(\sep)_{\nu}{}^{b}(\sep)_{\rho}{}^{c}\pd_{(a}\Lam_{bc)}:=0$. This is true
even if the parameter $\Lam_{ab}$ had contained a trace which could play the role of a longitudinal \scale3 transformation. This is interesting since the Einstein tensor is not invariant under general \scale3 transformations. There is in fact a two-derivative expression, the ``Ricci'' tensor, that is also \transl3 invariant but only for trace-free parameters as noticed already in \cite{Damour:1987vm}. Its relation to $\tom_{\mu}{}^{ab}|_{\mr{lin}}$ is analyzed below.

Following  the linearized analysis of 
\cite{Bergshoeff:2009tb} the \scale3 invariant Cotton tensor\footnote{For the precise statement about spin 2 scale invariance see the discussion at the very end of section 2.} can now be  constructed in two different ways from the Einstein tensor. One is, dropping the $|_{\mr{lin}}$ on the linearized quantities in the remainder of this subsection,
\beq
C_{\mu\nu\rho}=\thalf\Box G_{\mu\nu\rho}-\tfrac{3}{8}(\eta_{(\mu\nu}\Box-\pd_{(\mu}\pd_{\nu})G_{\rho)\si}{}^{\si},
\eeq
which is totally symmetric, but this is imposed by hand.
A second, perhaps more convenient,  way to write the Cotton tensor is in terms of the spin 3 Schouten tensor in (\ref{schoutenthree}) as
\beq
\label{cottonthreemetric}
C_{\mu\nu\rho}:=(\sep)_{\mu}{}^{\si}(\sep)_{\nu}{}^{\tau}S_{\si\tau\rho},
\eeq
which is, as noted in \cite{Henneaux:2015cda}, automatically symmetric and hence also trace and divergence free on all indices. This last way to write the Cotton tensor is, as we will see below, more directly related to the frame field formulation and the  cascade equations in our gauge. This fact becomes clear if one notices that the cascade field $\tf_{\mu}{}^{ab}$  in the metric formulation can be made to correspond to the Schouten tensor. Utilizing this fact we will towards the end of this subsection provide an exact relation between
the frame field $e_{\mu}{}^{ab}$ and the ``metric'' $h_{\mu\nu\rho}$.

In \cite{Damour:1987vm} the spin 3 Schouten tensor is defined in analogy with the spin 2 case, namely as the part (in any dimension) of
the Riemann tensor left over when subtracting the Weyl tensor. Of course, in three dimensions and for any spin, the Weyl tensor is zero (as shown in appendix \ref{app:A}
 for spin 2 but this proof can be applied also to higher spins)
and the Riemann tensor is given entirely in terms of the Schouten tensor. The relation in three dimensions for spin 3 is given in \cite{Damour:1987vm} (with their conventions but writing out the  antisymmetrizations explicitly)
\beq
R_{\mu\nu\rho,\al\be\ga}=\pd_{\mu}\pd_{\nu}\pd_{\rho}h_{\al\be\ga}|_{[\mu\al],[\nu\be],[\rho\ga]}=\tfrac{1}{3}(S_{\mu\nu\al\ga}\eta_{\rho\be}|_{[\al\rho],[\ga\be]}+\cdots),
\eeq
where the ``Schouten tensor'' 
\beq
S_{\mu\nu\al\be}:=R_{\mu\nu\al\be}-\tfrac{1}{16}R_{\mu(\nu}\eta_{\al\be)}|_{[\mu\nu]}.
\eeq
Here the authors of \cite{Damour:1987vm} have used the definitions $R_{\mu\nu\al\be}=\pd_{[\mu}R_{\nu]\al\be}$ and $R_{\mu\nu}=\pd_{[\mu}R_{\nu]\al}{}^{\al}$ in terms of the 
``Ricci'' tensor 
\beq
\label{ricci3}
R_{\mu\nu\rho}:=\Box h_{\mu\nu\rho}-\pd^{\al}\pd_{(\mu}h_{\nu\rho)\al}+\pd_{(\mu}\pd_{\nu}h_{\rho)\al}{}^{\al}. 
\eeq
Using the connection to the frame field formulation established below we will be able to conclude
that this Ricci tensor is closely related to the metric form of the cascade field $\tom_{\mu}{}^{ab}$, both of which are \transl3 invariant for trace-free parameters only.

Using the above  definitions quoted from the  previous works \cite{Damour:1987vm} and \cite{Bergshoeff:2009tb} 
we can  now nail down the connection between our basic frame field and the ``metric''. Let us first define a completely symmetric field $h_{\mu\nu\rho}$
by its relation to the frame field $e_{\mu}{}^{ab}$ (traceless in $ab$) as follows
\beq
e_{\mu}{}^{ab}=h_{\mu}{}^{ab}+c\,e_{\mu}{}^{(a}h^{b)}-\tfrac{1}{3}(c+1)h_{\mu}\eta^{ab},
\eeq
where the trace $h_{\mu\nu}{}^{\nu}:=h_{\mu}$. 
Hence, requiring just tracelessness in $ab$ of the RHS in this relation  leaves a free parameter $c$.
Using this equation one can check that the linearized Cotton tensor in the frame field formulation as given in (\ref{framecotton}) produces precisely the Cotton tensor of ref. \cite{Bergshoeff:2009tb} quoted above. The free parameter $c$ is thus not determined at this stage which is not surprising in view of the fact that the Cotton tensor is \scale3 invariant.
We thus expect that an identification between  some other fields in the metric and frame field formulations  will be required to determine $c$.

For the simple relation  $h_{\mu\nu\rho}:=e_{(\mu\nu\rho)}$ to be true we need 
actually $c=\thalf$ and then
\beq
\label{threemetric}
e_{\mu}{}^{ab}=h_{\mu}{}^{ab}+\thalf(e_{\mu}{}^{(a}h^{b)}-h_{\mu}\eta^{ab}).
\eeq
Indeed, the  value $c=\thalf$ is obtained by identifying the linearized  and symmetrized three derivative cascade field $\tf_{\mu}{}^{ab}$ with the
Schouten tensor in (\ref{schoutenthree}), namely the one 
constructed in terms of the ``metric'' $h_{\mu\nu\rho}$ in \cite{Damour:1987vm}  and \cite{Bergshoeff:2009tb}. Note that neither the Schouten tensor nor the (by hand) symmetrized cascade field $\tf_{\mu}{}^{ab}$ is
trace free (in all index pairs) and the identification works also for these irreducible trace parts\footnote{In \cite{Nilsson2013} the cascade field $f_{\mu}{}^{ab}$ was given the name ``Schouten tensor'' which in view of the present discussion is more appropriately associated to $\tf_{\mu}{}^{ab}$.}. In order to get a better understanding of the connection between the frame field and the metric, we conduct in the next subsection  a more thorough discussion of the implementation of the spin 3 metric gauge using \lorentz3. This will include
a more  general comparison of linear as well as non-linear properties of the relevant symmetries.

\subsection{Symmetries: linear vs. non-linear}

We will here discuss how the different cascade fields behave under the various spin 3 symmetries and to what extent this affects the implementation of various gauges, the main one being the metric gauge. In doing so it will also be interesting  to compare the linearized to the full non-linear results. Why this is so  will be clear below.

As a first step we would like to gauge fix the spin 3 frame field $e_{\mu}{}^{ab}$ (containing irreps 
${\bf 7}\oplus{\bf 5}\oplus {\bf 3}$) to the corresponding ``metric'' $h_{\mu\nu\rho}$\,, which is totally symmetric in all three indices and has a non-zero trace. This can be accomplished using the spin 3 Lorentz symmetry \lorentz3 with parameter $\Lam^{ab}(3,1)$
and is easily done by setting $\de e_{\mu}{}^{ab}|_{\bf 5}=0$. In fact, dropping the spin 2 transformation terms and implementing the spin 2 gauge $b_{\mu}=0$  (see appendix \ref{app:C}) in
\beq
\de e_{\mu}{}^{ab}(4,0)=\mr{D}_{\mu}\Lam^{ab}(4,0)+\ep_{\mu}{}^{d(a}\Lam^{b)}{}_d(3,1)-e_{\mu}{}^{(a}\Lam^{b)}(3,1),
\eeq
we can  solve for $\Lam^{ab}(3,1)$. Inserting this solution back into the equation for $\de e_{\mu}{}^{ab}$ just projects it down to the sum of irreps ${\bf 7}\oplus {\bf 3}$. This has immediate  implications for how to use  the relation $e_{\mu ab}=h_{\mu ab}+\thalf(\eta_{\mu (a}h_{b)}-h_{\mu}\eta_{ab})$ when deriving the transformation rules for the spin 3 metric $h_{\mu ab}$. Explicitly, the solution for the \lorentz3 
parameters is
at the linearized level
\beq
\Lam^{ab}(3,1)=-\tfrac{2}{3}\ep^{\mu\nu (a}\pd_{\mu}\Lam_{\nu}{}^{b)}(4,0).
\eeq
Thus we see that  when compensating the \transl3 by a Lorentz transformation with this parameter the result is (neglecting here also the \scale3 part of the transformation)
\beq
\de e_{abc}=\pd_{a}\Lam_{bc}(4,0)+\ep_{ad (b}\Lam_{c)}{}^d(3,1)=\pd_{(a}\Lam_{bc)}(4,0)+\tfrac{1}{3}(\eta_{a(b}\pd^d\Lam_{c)d}(4,0)-
\eta_{bc}\pd^d\Lam_{ad}(4,0)).
\eeq
Using now the transformation rule $\de h_{abc}=\pd_{(a}\Lam_{bc)}(4,0)$ (and hence $\de h_{a}=\tfrac{2}{3}\pd^{b}\Lam_{ab}(4,0)$) we find, as expected, that
\beq
\de e_{abc}=\de h_{abc}+\thalf(\eta_{a (b}\de h_{c)}-\de h_{a}\eta_{bc}).
\eeq

Having established the relation between the frame field and the metric in the spin 3 sector we can check how the cascade fields as functions of the metric behave under various symmetries. First we consider $\te_{\mu}{}^{ab}(e)|_{\mr{lin}}$. This cascade field  is manifestly invariant under \transl3 while as a function of the metric ($\te_{\mu}{}^{ab}(h)|_{\mr{lin}}$) it is not. In fact, this is obvious already for 
$(\sep e)_{\mu}{}^{ab}$, which is \transl3 invariant, while $(\sep (\de h))_{\mu}{}^{ab}\neq 0$.

Turning to  the next cascade field, $\tom_{\mu}{}^{ab}|_{\mr{lin}}$, let's consider it too as a function of the metric. The symmetrized expression then reads
\beq
\label{linomegatildemetric}
\tom_{(abc)}(h)|_{\mr{lin}}=-\thalf(\Box h_{abc}-3\pd_{(a}\pd^{\mu}h_{bc)\mu}+3\pd_{(a}\pd_bh_{c)}
+\tfrac{1}{3}\eta_{(ab}(2\pd^{\mu}\pd^{\nu}h_{c)\mu\nu}-2\Box h_{c)}-\pd_{c)}\pd^{\mu}h_{\mu})),
\eeq
where the first three terms are the ones used by Damour and Deser in \cite{Damour:1987vm} 
in their definition of the ``Ricci'' tensor given in eq. (\ref{ricci3}) above\footnote{Note the different normalization of symmetry brackets as compared to \cite{Damour:1987vm} (see eq. (\ref{ricci3})).}. This set of terms is  separately invariant under linearized \transl3 transformations, as are the remaining terms. 

One should perhaps note in this context that in the frame formulation $\tom_{\mu}{}^{ab}(e)|_{\mr{lin}}$ is of course \transl3 invariant since this is true already for $\te_{\mu}{}^{ab}(e)|_{\mr{lin}}$.\footnote{Note that since (in our spin 3 guage) $\tom_{\mu}{}^{ab}(e)|_{\mr{lin}}$ comes from squaring ${\mathcal O}=\sep-2\sE$ there are (at least) four expressions bilinear in derivatives that are separately invariant. Here we use the fact that given the operators ${\mathcal O}$ and $\sep$ defined above this equation defines $\sE$. This follows directly from the fact that
already the first order expressions $(\sep e)_{\mu}{}^{ab}$ and $(\sE e)_{\mu}{}^{ab}$ appearing in  $\te_{\mu}{}^{ab}$ as a function of $e_{\mu}{}^{ab}$ are \transl3 invariant.} However, this is no longer true  for $\te_{\mu}{}^{ab}$ if it is written in terms of the ``metric'' $h_{abc}$ as already mentioned above. The logic here is that for a \transl3 invariant cascade field expressed in terms of the frame field $e_{\mu}{}^{ab}$ to stay  invariant when the \lorentz3 invariance is used to turn the frame field into the ``metric'' it must itself be \lorentz3 invariant: this is true for  $\tom_{\mu}{}^{ab}(e)|_{\mr{lin}}$ but not for $\te_{\mu}{}^{ab}(e)|_{\mr{lin}}$ (see appendix \ref{app:C}).
It is interesting to note here that at the non-linear level  $\tom_{\mu}{}^{ab}(e)$ is no longer invariant
under the spin 3 $\tLam^{ab}(3,1)$ Lorentz transformations and hence $\de_{\text{\transl3}}\tom_{\mu}{}^{ab}(h)\neq 0$ as we have also verified by an explicit Mathematica calculation.

A similar discussion for the \scale3 transformations gives a completely different result.
In fact, from appendix \ref{app:C} we have, dropping the spin 2 parameters,
\beq
\de \te_{\mu}{}^{ab}=\mr{D}_{\mu}\Lam^{ab}(3,1) -2\ep_{\mu}{}^{d(a}\Lam^{b)}{}_d(2,2)-e_{\mu}{}^{(a}\Lam^{b)}(2,2)-4\ep^{cd(a}f_{\mu c}\Lam^{b)}{}_d(4,0),
\eeq
and
\beq
\de \tom_{\mu}{}^{ab}=\mr{D}_{\mu}\Lam^{ab}(2,2) 
+3\ep_{\mu}{}^{d(a}\Lam^{b)}{}_d(1,3)-e_{\mu}{}^{(a}\Lam^{b)}(1,3)
+3\ep^{cd(a}f_{\mu c}\Lam^{b)}{}_d(3,1)+f_{\mu}{}^{(a}\Lam^{b)}(3,1).
\eeq
Then using the gauge condition  $\de\te_{\mu}{}^{a}(3,1)=0$ we can solve for some of the gauge parameters that appear on the RHS of the last two equations above:
\beq
\Lam^{ab}(2,2)=D^{(a}\Lam^{b)}(3,1)+6f^{(a}{}_c\Lam^{b)c}(4,0)-\trace,
\eeq
and
\beq
\Lam^{a}(2,2)=-\tfrac{1}{3} \ep^{abc} \mr{D}_b\Lam_c(3,1)-2\ep^{abc}f_{bd}\Lam_c{}^d(4,0).
\eeq
From these expressions we conclude that  linearly $\tom_{\mu}{}^{ab}(e)$ transforms under \scale3 but not under either
\transl3 or \lorentz3. At the non-linear level all three transformations affect $\tom_{\mu}{}^{ab}(e)$. We now turn to a more systematic discussion of the non-linear properties of the spin 3 system and in particular the non-linear properties of the Cotton equation.

In addition to the results quoted above for the $\Lam^{(a)}(2,2)$, the Stückelberg gauges lead to the following spin 3 parameter relations
\beq
\Lam^{ab}(0,4)=\tfrac{1}{6}(D^{(a}\Lam^{b)}(1,3)+
f^{(a}{}_c\Lam^{b)c}(2,2)+\tfrac{3}{2}\ep^{cd(a}f^{b)}{}_{c}\Lam_d(2,2)+ 2f^{ab}\Lam(2,2)-\trace),
\eeq
\beq
\Lam^{ab}(1,3)=\tfrac{1}{3}(D^{(a}\Lam^{b)}(2,2)+
3f^{(a}{}_c\Lam^{b)c}(3,1)-3\ep^{cd(a}f^{b)}{}_{c}\Lam_d(3,1)-\trace),
\eeq
\beq
\Lam^a(1,3)=\tfrac{1}{6}(\ep^{a\mu\nu}\mr{D}_{\mu}\Lam_{\nu}(2,2)+3\ep^{a\mu\nu}f_{\mu}{}^a
\Lam_{\nu a}(3,1)+3f^{ab}\Lam_b(3,1)-3f_b{}^b\Lam^a(3,1),
\eeq
and
\beq
\Lam(2,2)=\tfrac{1}{6}\mr{D}_{\mu}\Lam^{\mu}(3,1)+f_{ab}\Lam^{ab}(4,0).
\eeq
One observation that can be made in relation to these expressions is that at the linear level (i.e., setting $e_{\mu}{}^a=\de_{\mu}^a$) only parameters with one index appear on the RHSs 
and hence they are all determined by the dilatation parameter $\Lam^a(3,1)$. Hence the only dependence of the transformation rules for the cascade fields on the spin 3 translations and Lorentz symmetries come from the explicit dependence in the transformation rules (see appendix \ref{app:C}). 

We conclude that at the linear level the metric gauge can be imposed without affecting the translation invariance for the cascade fields $\tom^{ab}$, $\tf^{ab}$ and  $f^{ab}$ but not for $\te^{ab}$. In the case of \scale3 invariance the first quantity that can be gauge fixed is the Cotton tensor as is clear already at linear level.

This should be compared to the completely different  situation that is at hand at the non-linear level. Here the above expressions show that new dependencies of $\de(cascade)$ on all parameters $\Lam^{ab}(4,0)$, $\Lam^{ab}(3,1)$ and $\Lam^{a}(3,1)$
are induced after implementing the Stückelberg gauges. Thus the first non-linear quantity to be invariant under all symmetries is the Cotton tensor since it is a component of the field strength $F_3$. This  probably means that
the task to construct a non-linear theory starting from the linear metric formulation
is more complicated than what one might have  anticipated.

\subsection{The non-linear Cotton equation}
\label{sec:nonlinCotton3}
We have now come to a point where the full non-linear structure of the spin 3 Cotton equation, covariantly coupled to spin 2, can be discussed and explained. 
We emphasize again that we only deal with the terms in this equation which originate from single commutators (i.e., the first term) in the expansion of the star product or, which in this particular case has the same affect, setting all fields with $s\geq 4$ to zero\footnote{The corresponding statement is not true for the spin 4 and higher Cotton equations.}.

In terms of the HS gauge system the (two-form) spin 3 Cotton equation is 
\beq
F^{ab}(0,4)=0,
\eeq
which reads, after dualizing and using the HS algebra but before any substitutions from the cascade solution have been done,
\begin{multline}
({\smcE}f)_{\mu}{}^{ab}+\ep_{\mu}{}^{c(a}f_c{}^{b)}\hat f-(f^{ab}\hat f_{\mu}-\tfrac{1}{3}\eta^{ab}f\hat f_{\mu})+(\de_{\mu}^{(a}f^{b)c}\hat f_c-
\tfrac{1}{3}\eta^{ab}f_{\mu}{}^c\hat f_c)\\
-f_{\mu}{}^c\tf^{(ab)}{}_c+f^{c(a}\tf_c{}^{b)}{}_{\mu}+\de_{\mu}^{(a}f\tf_c{}^{b)c}-f\tf^{(ab)}{}_{\mu}-\de_{\mu}^{(a}f_{de}\tf^{|de|b)}-f_{\mu}{}^{(a}\tf_c{}^{b)c}=0,
\end{multline}
where ${\smcE}$  is the spin 2 covariantized version of ${\sep}$
defined in the first subsection (where also its action on any tensor of the type $e_{\mu}{}^{ab}(4,0)$ was given). In the above Cotton equation we have also defined  $f:=f_a{}^a(0,2)$.
Substituting the cascade solution into this Cotton equation to express it in terms of only the frame field $e_{\mu}{}^{ab}$  gives rise to an equation far too big to write out in detail here. 
The equation will therefore be presented in an exact but unrefined form of the appended raw output \cite{LinanderMath} from our Mathematica program. This version is not processed beyond basic canonicalization and collection of explicit index symmetries, i.e. keeping the structure from the cascade solution\footnote{Although other, in some sense more simplified, forms have been useful in the process of verifying its properties, this is still one of the most compact expressions. For example, performing the explicit commutators in \cite{LinanderMath} tends to increase the total number of terms (after canonicalization). }. 

One way to present the complete Cotton equation is, as mentioned in the beginning of this section, as the sum of two parts one containing  all terms with five explicit
derivatives and one containing all terms with one or two spin 2 Schouten tensors (and thus less than five explicit derivatives)\footnote{As mentioned previously, all terms have five derivatives if we also count the two inside the spin 2 Schouten tensor.}. The first part was discussed at the linearized level in the previous subsection and can be taken from there by replacing the linearized operators $\mathcal O$ and $\hat{\mathcal O}$ by their spin 2 covariantized analogues
which we denote $\mathcal D$ and $\hat{\mathcal D}$. Since $\mathcal O$ and $\hat{\mathcal O}$ do not commute (recall their action on $e_{\mu}{}^{ab}$ type tensors) there is no ordering problem associated with this procedure as there generally is when trying to \lorentz2 covariantize  expressions written in terms of partial derivatives.

As an illustration of this procedure we present the cascade field $\tom_{\mu}{}^{ab}$ this way:
\beq
\tom_{\mu}{}^{ab}=\tom_{\mu}{}^{ab}(D)+\tom_{\mu}{}^{ab}(f),
\eeq
where the two-derivative part is
\beq
\tom_{\mu}{}^{ab}(D)=-\thalf({\mathcal D}^2\,e)_{\mu}{}^{ab},
\eeq
which when linearized reduces to the expression in eq. (\ref{tomlinear}), while
 the no-derivative part reads (here just the {\it f(spin 2)e(spin 3)} part of the solution for $\tom_{\mu}{}^{ab}$ given in section \ref{sec:solution})
\begin{multline}
\tom_{\mu}{}^{ab}(f)=2(2f_{[\mu}{}^{\nu}e_{\nu]}{}^{ab}+f^{c(a}e_{\mu c}{}^{b)}-f_{\mu}{}^{\nu}e^{(ab)}{}_{\nu}+f^{c(a}e^{b)}{}_{c\mu}-f^{(ab)}e_{c\mu}{}^c)\\
-\tfrac{4}{3}\eta^{ab}(f_{[\mu}{}^ce_{d]}{}^d{}_c-f_{[c}{}^de_{d]}{}^c{}_{\mu}),
\end{multline}
where we have also implemented the fact that the spin 2 Schouten tensor $f_{\mu\nu}(0,2)$ is symmetric. 
Recalling the linearized expression in (\ref{tomlinear}) one realizes  that even in this trivial case it would be   quite  tedious to  compute the first part $\tom_{\mu}{}^{ab}(D)=-\thalf({\mathcal D}^2\,e)_{\mu}{}^{ab}$ explicitly. However, if this is done one gets further contributions to $\tom_{\mu}{}^{ab}(f)$ but their precise form  depends on how one chooses to define the (non-zero) two-derivative term. This is of course true for all 
non-linear expressions considered in this paper.

We now turn to the expression for the full non-linear Cotton tensor. Written as the sum of the two terms as explained above it reads
\beq
C_{\mu}{}^{ab}=C_{\mu}{}^{ab}(D)+C_{\mu}{}^{ab}(f),
\eeq
where the first term 
\beq
C_{\mu}{}^{ab}(D)=\tfrac{1}{4!}{\smcE}(\mcD^4e+\mcD\mchD\mcD^2e)_{\mu}{}^{ab},
\eeq
whose explicit form is best obtained using the computer while the second one splits naturally into
\beq
C_{\mu}{}^{ab}(f)=C_{\mu}{}^{ab}(f,D^3)+C_{\mu}{}^{ab}(f^2,D),
\eeq
where the notation on the RHS (the comma) indicates that the derivatives can also act on the spin two Schouten tensor  $f$.
These latter two parts of the spin 3 Cotton equation are also too complicated to be given here (however, see \cite{LinanderMath}). We hope to present a more manageable expression
for the this spin 3 Cotton equation elsewhere
\cite{LinanderNilsson}.

Apart from this we should also check that the necessary symmetries are present. The \transl3 and \scale3 transformations are checked using our Mathematica based system
and both are found to work as expected: all terms with derivatives on the gauge parameters cancel and the remaining terms appear in the following form
\beq
\de F(0,4)=[F(1,1), \Lam(0,4)]+[F(0,2), \Lam(1,3)],
\eeq
where both $\Lam^{ab}(0,4)$, $\Lam^{ab}(1,3)$ and $\Lam^{a}(1,3)$ can at the non-linear level be related to $\Lam^{ab}(4,0)$, $\Lam^{ab}(3,1)$ and $\Lam^{ab}(3,1)$ by cascading (as explained previously in this section). This then implies that {\it invariance} under all three (including \lorentz3) of these symmetries arise only modulo the spin 2 Cotton equation $F(0,2)=0$. Note that also $F(1,1)$ appears in $\de F(0,4)$ above but $F(1,1)=0$ has been solved and will therefore not
arise in the computation of $\de F(0,4)$.

Thus due to these \lorentz3 properties, the metric formulation of the spin 3 Cotton equation is  obtained by simply inserting (\ref{threemetric}) into the frame field Cotton equation. It can be checked (using Mathematica) that the spin 3 Cotton tensor is totally symmetric and hence traceless and divergencefree in all indices as expected. This answers the questions i) - iii) posed in  subsection 4.1. 

Perhaps surprisingly,  we find that the invariance under \transl3 does not arise until one reaches the Cotton equation in the cascade  while for the \scale3 invariance this is
 a well-known fact already from the linearized analysis. In other words, the fact 
that there exists a linearized expression with two partial derivatives acting on the spin 3 metric  that is  invariant under \transl3 (true also for the frame field formulation above) does not continue to hold
at the full non-linear level. It is not clear to us if this could have been foreseen without consulting the Chern-Simons formulation.
This property continues to be true when considering the next field in the cascade $\tf_{\mu}{}^{ab}$ corresponding  to the Schouten tensor which may be even more surprising.

\section{Aspects of the spin 4 sector}

The theory based on the star product was constructed by Fradkin and Linetsky some time ago in \cite{Fradkin:1989xt} where also the supersymmetric version was considered. In that paper the bosonic theory analyzed here was written out with all spins from two  to infinity appearing explicitly in a Lagrangian including  all kinetic and cubic interaction terms. However, their results come directly from the Chern-Simons gauge theory and appears in \cite{Fradkin:1989xt} in the version with the auxiliary fields, i.e., before the elimination of the Stückelberg and the dependent fields that we have performed in this paper. The higher order interaction terms can thus not be read off from their form of the Lagrangian. The structure of the cubic terms in this Lagrangian, with the rather complicated coefficients, corresponds to computing the HS multicommutators for the quadratic terms in the field equations or, alternatively, deducing them by Taylor expanding the star product. In writing out the  Chern-Simons Lagrangian for all spins also 
a formula for the  trace in the HS algebra is needed, see \cite{Fradkin:1989xt}.

The expanded version of the theory in \cite{Fradkin:1989xt} thus provides an explicit form of the
Lagrangian. However, if one is interested in the general structure of interaction terms between three or more frame fields with given spins
 one has to eliminate all auxiliary fields, Stückelberg as well as dependent ones, in order to
reduce the field content down to just the basic spin $s\geq 2$ frame fields $e_{\mu}{}^{a_1...a_{s-1}}$. Unfortunately, at the non-linear level this is an arduous
task  which quickly makes the use of computer methods unavoidable. This is amply demonstrated already by the spin 3 equations discussed above in this paper. 

So far in this paper we have focused on the spin 2 - spin 3 subsystem  to develop techniques (in particular for the computer) and 
a better understanding of the general structure. The complete HS theory with all integer spins $s\geq 2$
is, however, reducible in the sense that it can be consistently truncated to contain only the even spins.
This follows directly form the algebra which has the property that two generators both with odd or even spins
(multi)commute, in either case, to generators with odd spin while a pair of generators, one with even and one with odd spin, has (multi)commutators that contain only odd spins. For the frame fields themselves (with spin $s(\mr{field})=j(\mr{generator}) + 1$ due to the one-form index) this translates into the result stated in section 2.2 
for the consistency of the even spin truncation. 

As  a first step towards addressing the more intricate aspects of the theory of (only) even spins we here analyze the structure of the component content of the equations $F_4=0$ for spin 4 to extract the constraints and the cascade equations, now setting all fields with  spins  higher than four to zero. Again we follow closely the procedure outlined in \cite{Pope-Townsend89}. As we will see some understanding gained in the study of the
spin 3 case can be immediately taken over to the spin 4 case. For instance, the feature  that
in the star product theory the spin two field equation  picks up interactions terms with two  spin 4 fields from the third order commutator is true also for  two  spin 3 fields but in this latter case  these terms are neglected
in the single commutator analysis carried out in this paper. The star product features of the interaction terms can, on the other hand, be read off from the following list of low spin star product commutators
(for spin $s$ fields $A_s$ valued in the spin $j=s-1$ part of the HS star algebra)
 \beqa
 \label{fieldinteractions}
 [A_2,A_2]_*&=&A_2,\cr
  [A_2,A_3]_*&=&A_3,\cr
   [A_3,A_3]_*&=&A_4+A_2,\cr
    [A_2,A_4]_*&=&A_4,\cr
     [A_3,A_4]_*&=&A_5+A_3,\cr
      [A_4,A_4]_*&=&A_6+A_4+A_2,\cr
       [A_2,A_5]_*&=&A_5\,,
 \eeqa
 where the first column of terms on the RHSs come from single commutators (or Poisson brackets in this paper), the second column from third order commutators, etc, in the expansion of the star product commutator. Here we have not included the abelian vector field $A_1$ which takes values in the central charge element $G(0,0)$ which is independent of the operators $q^{\al}$ and $p_{\al}$.

The expansion of the spin 4 gauge one-form potential reads
\beqa
\label{gaugefieldspinfour}
A_4&=&e^{abc}(6,0)P_{abc}(6,0)+\cr
&&\te^{abc}(5,1)\tP_{abc}(5,1)+\te^{ab}(5,1)\tP_{ab}(5,1)+\cr
&&\te^{abc}(4,2)\tP_{abc}(4,2)+\te^{ab}(4,2)\tP_{ab}(4,2)+\te^{a}(4,2)\tP_{a}(4,2)+\cr
&&\tom^{abc}(3,3)\tM_{abc}(3,3)+\tom^{ab}(3,3)\tM_{ab}(3,3)+\tom^{a}(3,3)\tM_{a}(3,3)+\tb(3,3)\tD(3,3)+\cr 
&&\tf^{abc}(2,4)\tK_{abc}(2,4)+\tf^{ab}(2,4)\tK_{ab}(2,4)+\tf^{a}(2,4)\tK_{a}(2,4)+\cr
&&\tf^{abc}(1,5)\tK_{abc}(1,5)+\tf^{ab}(1,5)\tK_{ab}(1,5)+\cr
&&f^{abc}(0,6)K_{abc}(0,6),
\eeqa
where we now keep the $(n_q,n_p)$ notation for both fields  and generators in order 
to avoid cluttering the symbols for these quantities with more things than tildes\footnote{We still use tilde to indicate that the flat indices are irreps when this is not automatic.}. Some of the fields in $A_4$ with less than three flat indices are Stückelberg and can be gauged away. However, as seen before 
in this paper this gauging fixing procedure is not unique and we choose here to eliminate fields in the manner indicated in table 4 by the left pointing arrows.
\begin{table}
\centering
  \renewcommand*{\arraystretch}{1.3}
  \[ 
  \begin{array}{l@{\hskip 0.3cm}c@{\hskip 0.3cm}llcl}
                               \multicolumn{2}{r}{\text{Frame field: }}            &e_{\mu}{}^{abc}(6,0)       &\\
                            \hline
F_{\mu\nu}^{abc}(6,0)=0     &\overset{\text{\small{solve}}}{\longrightarrow}&  \te_{\mu}{}^{abc}(5,1)   &&\\
& &\te_{\mu}{}^{ab}(5,1)=0  &               &\longleftarrow             & \Lam^{abc}(4,2), \Lam^{ab}(4,2), \Lam^{a}(4,2)\\
F_{\mu\nu}^{abc}(5,1)=0     &\overset{\text{\small{solve}}}{\longrightarrow}&  \te_{\mu}{}^{abc}(4,2)   &&\\
F_{\mu\nu}^{ab}(5,1)=0      &\overset{\text{\small{solve}}}{\longrightarrow}&  \te_{\mu}{}^{ab}(4,2)    &&\longleftarrow& \Lam^{abc}(3,3)\\
& &\te_{\mu}{}^{a}(4,2)=0   &               &\longleftarrow             & \Lam^{ab}(3,3), \Lam^{a}(3,3), \Lam(3,3)\\ 
F_{\mu\nu}^{abc}(4,2)=0     &\overset{\text{\small{solve}}}{\longrightarrow}&  \tom_{\mu}{}^{abc}(3,3)  &&\\
F_{\mu\nu}^{ab}(4,2)=0      &\overset{\text{\small{solve}}}{\longrightarrow}&  \tom_{\mu}{}^{ab}(3,3)   &&\longleftarrow& \Lam^{abc}(2,4)\\
F_{\mu\nu}^{a}(4,2)=0       &\overset{\text{\small{solve}}}{\longrightarrow}&  \tom_{\mu}{}^{a}(3,3)    &&\longleftarrow& \Lam^{ab}(2,4)\\
& & \tb_{\mu}(3,3)=0        &               &\longleftarrow             & \Lam^{a}(2,4)\\
F_{\mu\nu}^{abc}(3,3)=0     &\overset{\text{\small{solve}}}{\longrightarrow}&  \tf_{\mu}{}^{abc}(2,4)   &&\\
F_{\mu\nu}^{ab}(3,3)=0      &\overset{\text{\small{solve}}}{\longrightarrow}&  \tf_{\mu}{}^{ab}(2,4)    &&\longleftarrow& \Lam^{abc}(1,5)\\
F_{\mu\nu}^{a}(3,3)=0       &\overset{\text{\small{solve}}}{\longrightarrow}&  \tf_{\mu}{}^{a}(2,4)     &&\longleftarrow& \Lam^{ab}(1,5)\\
F_{\mu\nu}(3,3)=0           &               && \multicolumn{3}{l}{\text{\small nothing to solve for}}\\
F_{\mu\nu}^{abc}(2,4)=0     &\overset{\text{\small{solve}}}{\longrightarrow}&  \tf_{\mu}{}^{abc}(1,5)   &&\\
F_{\mu\nu}^{ab}(2,4)=0      &\overset{\text{\small{solve}}}{\longrightarrow}&  \tf_{\mu}{}^{ab}(1,5)    &&\longleftarrow& \Lam^{abc}(0,6)\\
F_{\mu\nu}^a(2,4)=0         &               && \multicolumn{3}{l}{\text{\small nothing to solve for}}\\
F_{\mu\nu}^{abc}(1,5)=0     &\overset{\text{\small{solve}}}{\longrightarrow}&  f_{\mu}{}^{abc}(0,6)     &&\\
F_{\mu\nu}^{ab}(1,5)=0      &               && \multicolumn{3}{l}{\text{\small nothing to solve for}}\\
\hline
F_{\mu\nu}^{abc}(0,6)=0     &\longrightarrow& \multicolumn{3}{l}{\text{Cotton equation.}}   
\end{array}
\]
\caption{Spin 4 system. The table indicates which field is expressed in terms of other fields by solving a specific $F=0$ component equation. It is also shown how, in this particular gauge, the gauge parameters are used to set to zero some or all irreps in the field components of $A_4$.}
\label{tab:spin4}
\end{table}
From table 4   we can also conclude that the following 13 equations actually constitute constraints (from lines having either both left and right  arrows or having no arrows):

\begin{equation}
\begin{array}{ccccc}
F_{\mu\nu}^{abc}(5,1)|_{\bf 7}=0,& F_{\mu\nu}^{ab}(4,2)|_{\bf 7}=0,& F_{\mu\nu}^{a}(4,2)|_{\bf 5}=0,&
F_{\mu\nu}^{a}(3,3)|_{\bf 7}=0,&F_{\mu\nu}^{a}(3,3)|_{\bf 5}=0,\\
F_{\mu\nu}^{a}(3,3)|_{\bf 3}=0,&F_{\mu\nu}^{ab}(2,4)|_{\bf 7}=0,&F_{\mu\nu}^{a}(2,4)|_{\bf 5}=0,&
F_{\mu\nu}^{a}(2,4)|_{\bf 3}=0,&F_{\mu\nu}^{a}(2,4)|_{\bf 1}=0,\\
F_{\mu\nu}^{ab}(1,5)|_{\bf 7}=0,&F_{\mu\nu}^{ab}(1,5)|_{\bf 5}=0,&F_{\mu\nu}^{ab}(1,5)|_{\bf 3}=0.
\end{array}
\end{equation}

As for the lower spins discussed in previous sections of this paper, there are gauge symmetries not being utilized at the point of the analysis defined by table \ref{tab:spin4}: for spin 4 these are
\beq
\Lam^{abc}(6,0),\,\,\Lam^{abc}(5,1),\,\,\Lam^{ab}(5,1),
\eeq
which as usual correspond to the spin 4 versions of the ``translation'', ``Lorentz'' and ``scale'' transformations, respectively. After the elimination of the auxiliary fields, these transformations act on the only remaining independent field in the theory, the frame field $e_{\mu}{}^{abc}(6,0)$, which is in the representation 
${\bf 9}\oplus{\bf 7}\oplus {\bf 5}$,
in the expected fashion. This frame field can thus be gauge fixed, using $\Lam^{abc}(5,1)$,  to a totally symmetric spin 4  ``metric'' field $h_{\mu\nu\rho\si}$ containing just the irreps ${\bf 9}$ and ${\bf 5}$. If needed this procedure can be continued
 using $\Lam^{ab}(5,1)$  to obtain a traceless ``metric'' field  in the irrep ${\bf 9}$. The field 
 $h_{\mu\nu\rho\si}$ here called ``metric'' is just the spin 4 Fronsdal field defined to be totally symmetric and ``double trace-free''  corresponding to the Young tableau with four symmetrized boxes and the singlet removed giving the representation content ${\bf 14}={\bf 9}\oplus {\bf 5}$. The free conformal field theory of this spin 4 ``metric'' was also studied in  \cite{Bergshoeff:2011pm}, and more recently in \cite{Henneaux:2015cda} as part of a more general HS 
 analysis including   the linearized Cotton tensor.

\section{Conclusions}

The main objective of this paper has been to perform a  non-linear analysis of
  the 2+1 dimensional conformal higher spin (HS) system defined  in terms of a  Chern-Simons theory based on the HS version $\mr{SO}(3,2)$. This includes a derivation of the non-linear spin 3 Cotton equation so this work can be considered as a direct continuation of \cite{Nilsson2013, Nilsson2015} where the procedure was laid out and some initial results presented.
 As explained in these papers the crucial step  is to solve the cascade equations and insert the solution back into the spin 3 Cotton equation. It turns out that the computations that could be done by hand at the linear level (see \cite{Nilsson2013, Nilsson2015}) quickly become unmanageable and we have therefore been forced to develop computer methods to deal with these equations.
 
 The true structure of the interactions between the spin 2 and 3 frame fields does not reveal itself
 until all auxiliary fields have been eliminated. At the level of the Chern-Simons theory with all the auxiliary fields present there are of course only cubic terms in the Lagrangian, see, e.g., \cite{Fradkin:1989xt}. Higher order interactions arise when the Stückelberg and dependent fields are eliminated as indicated by the analysis at the linear level in \cite{Pope-Townsend89}. These interactions are therefore not directly tied to the HS algebra. 
 
 In order to carry out the computations and to get  exact non-linear results 
 we have found it necessary to restrict the theory  to the first term in expansion of the star commutator and set all fields with spin 4 and higher to zero. In this truncation scheme the spin 3 Cotton equations still contains more than $10^3$ terms (in the unrefined form given in \cite{LinanderMath} there are 1302 terms).
 
 Another goal of this paper was to derive the metric formulation from the theory expressed in terms of the frame field. As discussed in the main text it is straightforward to go to the metric formulation for the spin 3 (or any spin in fact) Cotton equation by imposing a Lorentz gauge (here sometimes called the metric gauge). This is possible, as discussed after eq. (\ref{linomegatildemetric}), since this equation is \lorentz3 invariant on-shell. At the linear level this step can also be taken for some of the cascade fields which can then be compared to various tensor fields  in the metric formulation already defined in the literature. In the spin 3 Stückelberg gauge employed in this paper (see table 1) the Schouten tensor $S_{\mu\nu\rho}(h)$ defined in (\ref{schoutenthree}) can be shown to be precisely our linearized cascade field 
 $\tf_{\mu}{}^{ab}(e)$ after converting the flat indices to curved ones and replacing the spin 3 frame field by the corresponding Fronsdal field using (\ref{threemetric}), i.e.,
 \beq
 e_{\mu}{}^{ab}=h_{\mu}{}^{ab}+\thalf(e_{\mu}{}^{(a}h^{b)}-h_{\mu}\eta^{ab}).
 \eeq
 The same statement is true also for the linearized spin 3 Cotton equation in terms of  (\ref{cottonthreemetric}) and 
 $F_{\mu\nu}{}^{ab}(0,4)(h)|_{\mr{lin}}=0$ after dualization.
 
 Using the metric definition (\ref{threemetric})
 we have also derived the exact non-linear metric form of the Schouten tensor and the Cotton equation which we believe have not appeared previously in the literature. Unfortunately, the spin 3 Cotton equation, in either the frame field or metric formulation, is too complicated to be presented explicitly. We hope to come back to this issue in a future publication. The full non-linear, but not completely refined, spin 3 Cotton equation in the frame field and metric  
 formulations are given in the appended files \cite{LinanderMath} and \cite{LinanderMath2}, respectively.

To extract information about this theory in $\mr{AdS}_3$ one could try to couple it 
to scalar fields as in \cite{Nilsson2015}.  The hope would then be that giving a VEV to some of these scalars the conformal symmetry will be spontaneously broken and the  theory develop an $\mr{AdS}_3$ background solution.
The prototypes here are the CFTs with six or eight supersymmetries coupled to superconformal Chern-Simons gravity \cite{Gran:2008qx, Chu:2009gi,   Gran:2012mg} where this is known to happen \cite{Chu:2009gi,Nilsson:2013fya}. However, if similar
phenomena can occur in the conformal HS theories coupled to scalars is far from clear. Of course, even without scalar fields
 one can just expand the conformal theory around a fixed $\mr{AdS}_3$ background and try to gain some  information about the  structure
of the higher derivative terms in the $AdS_3$ of the theory. Hopefully this may give some hints how to understand  the problematic issues of derivative dressing and locality that are known to be present in some
HS theories based on the Vasiliev construction.

In this paper we have also analyzed the linearized version of the spin 4 sector. For any spin we expect
there to be a natural definition of the Schouten tensor
in terms of the spin $s$ frame field analogues to the one in terms of the metric given in \cite{Henneaux:2015cda}.
Schematically we have for the Schouten and Cotton tensors
\beq
S(h)=\pd^{s}h,\,\,\,C(h)=\pd^{2s-1}h,
\eeq
which correspond in the frame formulation to,
respectively, the first cascade field after the ``spin connection'' $\tom(s-1,s-1)(e)=\pd^{s-1}e$ and the last component of the field strength $F$, i.e.,  
\beq
\tf_{\mu}{}^{a_1...a_{s-2}}(s-2,s)=\pd^se,\,\,\,F(0,2s-2)=\pd^{2s-1}e.
\eeq
Note, however, that the special form of the Schouten tensor used in previous works, i.e. (\ref{schoutenthree}), is also obtained in this paper but seems to be specific to the Stückelberg gauge used here. In, for instance, a gauge 
with the dilatation gauge fields $b_{\mu}$ set to zero
the relation to the standard metric Schouten tensor will be altered. The Cotton equation, on the other hand,  is of course not affected
by choosing different gauges.  

\section*{Acknowledgements}

We are very grateful to Nicolas Boulanger and Simone Giombi for discussions and comments 
on the paper prior to publication.

\appendix

\allowdisplaybreaks
\section{Conventions}
\label{app:A}

Tangent (flat) space indices are denoted by lower case Latin $a, b, \cdots$ while for curved space-time indices we use the second half of the lower case Greek alphabet $\mu, \nu, \cdots$. The Lorentzian metric $\eta^{a b}=diag(-1,+1,+1)$. The tangent space epsilon tensor is defined by $\epsilon^{0 1 2}=1$.

Spinor indices for the real $\mathrm{SL}(2, \mathbb{R})$ spinors use the first half of the lower case Greek alphabet $\alpha, \beta, \cdots$. They are raised and lowered from the left by the antisymmetric $\epsilon^{\alpha\beta}$ and $\epsilon_{\alpha\beta}$, respectively, obeying $\epsilon_{\alpha\beta}\epsilon^{\beta\gamma} = \delta_\alpha{}^\gamma$. The first (second) spinor index  on gamma matrices $(\ga^a)_{\al}{}^{\be}$ is raised and lowered from the left (right) which implies that $(\ga^a)_{\al}{}^{\be}p_{\be}=(\ga^a)_{\al}{}_{\be}p^{\be}$ but (note the minus sign) $p^{\al}(\ga^a)_{\al}{}^{\be}=-p_{\al}(\ga^a)^{\al}{}^{\be}$
which also define the symmetric gamma matrices with two upper or lower indices.

Sometimes we use the short hand $q\cdot p = q^\alpha p_\alpha$. The three dimensional gamma matrices connecting the spin and vector representation are chosen as
\begin{equation}
(\ga^a)_{\al}{}^{\be}:\,\,\,\,\,\gamma^0 = i \sigma^2\;,\quad \gamma^1 = \sigma^1 \;,\quad \gamma^2 = \sigma^3.
\end{equation}
\beq
\ga^0\ga^1\ga^2={\bf 1},\,\,\,\ga^{abc}=\ep^{abc}{\bf 1}\,.
\eeq

\subsection{Fierz identities}
\label{app:A1}

When calculating the commutation relations for the higher spin algebra some useful Fierz identities are

\beq
(\ga^a)_{(\al\be}(\ga_a)_{\ga)}{}^{\de}=0\,,
\eeq
\beq
(\ga^{[a})_{\al\be}(\ga^{b]})^{\ga\de}= \ep^{ab}{}_c(\ga^{c})_{(\al}{}^{(\ga}\de_{\be)}^{\de)}\,,
\eeq
\beq
(\ga^{[a})_{(\al\be}(\ga^{b]})_{\ga)}{}^{\de}=-\thalf \ep^{ab}{}_c(\ga^c)_{(\al\be}\de_{\ga)}^{\de}\,,
\eeq
\beq
(\ga^a)_{\al\be}(\ga_a)^{\ga\de}=2\de^{(\ga\de)}_{\al\be}=2(\ga^a)_{(\al}{}^{(\ga}(\ga_a)_{\be)}{}^{\de)}\,,
\eeq
\beq
(\ga^a)_{(\al}{}^{(\ga}(\ga^b)_{\be)}{}^{\de)}=\eta^{ab}\de^{(\ga\de)}_{\al\be}-(\ga^{(a}_{\al\be}(\ga^{b)})^{\ga\de}\,,
\eeq
\beq
(\ga^a)_{(\al(\ga}(\ga^b)_{\be)\de)}=\eta^{ab}\thalf(\ep_{\al\ga}\ep_{\be\de}+\ep_{\al\de}\ep_{\be\ga})-\ga^{(a}_{\al\be}\ga^{b)}_{\ga\de}\,.
\eeq

\subsection{Spin 2 curvature conventions}
\label{app:A2}

Here we summarize the conventions used in relating the usual Riemann tensor, Ricci tensor etc to the dualized Riemann tensor obtained from the dualized spin connection $\om_{\mu}{}^a$ used in section 2.

The Riemann tensor (and spin connection) is (right-)dualized as follows
\beq
R^{*}_{\mu\nu}{}^a=\thalf R_{\mu\nu}{}^{bc}\ep_{bc}{}^a,
\eeq
which is in this paper written without the asterisk. This tensor can be dualized a second time giving exactly the Einstein tensor
\beq
{}^*R^{*}_{\mu}{}^a=\tfrac{1}{4} \ep_{\mu}{}^{\rho\si}R_{\rho\si}{}^{bc}\ep_{bc}{}^a=R_{\mu}{}^a-\thalf e_{\mu}{}^a R.
\eeq

The relation between the Riemann tensor and the Schouten tensor $S_{\mu}{}^a$ is in 2+1 dimensions
\beq
R_{\mu\nu}{}^{ab}=4\de_{[\mu}^{[a}S_{\nu]}{}^{b]},
\eeq
where the Schouten tensor is here defined in terms of the Ricci tensor by
\beq
S_{\mu}{}^{a}=R_{\mu}{}^a-\tfrac{1}{4}e_{\mu}{}^a\,R.
\eeq
The fact the Weyl tensor is zero in $2+1$ dimensions is most easily shown by evaluating the expression on the LHS of the equation
\beq
\de_{[\mu\nu\rho\si]}^{abcd}W_{cd}{}^{\rho\si}=0.
\eeq
The tracelessness of the Weyl tensor then just means that the LHS is equal to $\tfrac{1}{6}W_{\mu\nu}{}^{ab}$.

\section{Relevant parts of the HS algebra}
\label{app:B}
The commutators involving the generators in the spin 2 and 3 sectors used in this paper are tabulated below.
\subsection{The spin 2 Poisson algebra}
\label{app:B1}

The algebra generated by the spin 2 generators $P^a(2,0)$, $M^a(1,1)$, $D(1,1)$ and  $K^a(0,2)$  is
\beqa
[M^a,M^b]&=&\ep^{ab}{}_cM^c,\cr
[M^a,P^b]&=&\ep^{ab}{}_cP^c,\cr
[M^a,K^b]&=&\ep^{ab}{}_cK^c,\cr
[P^a,K^b]&=&-2\ep^{ab}{}_cM^c-2\eta^{ab}D,\cr
[D, P^a]&=&P^a,\cr
[D, K^a]&=&-K^a,
\eeqa
where we have simplified the notation by dropping the $(n_q,n_p)$.

\subsection{The spin 2 - spin 3 Poisson brackets}
\label{app:B2}
\label{sec:hsalgebra}

The commutators below contain one generator from each of  the spin 2 and spin 3 sectors \cite{Nilsson2013}:
\beqa
[P^a(2,0),\;\tP^{bc}(3,1)]&=&\ep^{a(b}{}_dP^{c)d}(4,0),\cr
[P^a(2,0),\;\tP^b(3,1)]&=&-P^{ab}(4,0),\cr
[P^a(2,0),\;\tilde M^{bc}(2,2)]&=&-2\ep^{a(b}{}_d \tilde P^{c)d}(3,1)-(\eta^{a(b}\tilde P^{c)}(3,1)-\tfrac{1}{3}\eta^{bc}\tilde P^a(3,1)),\cr
[P^a(2,0),\;\tilde M^b(2,2)]&=&-\tilde P^{ab}(3,1)+\tfrac{3}{2}\ep^{ab}{}_c\tilde P^c(3,1),\cr
[P^a(2,0),\;\tilde D(2,2)]&=&-2\tilde P^a(3,1),\cr
[P^a(2,0),\;\tK^{bc}(1,3)]&=&3\ep^{a(b}{}_d\tilde M^{c)d}(2,2)-3(\eta^{a(b}\tilde M^{c)}(2,2)-\tfrac{1}{3}\eta^{bc}\tilde M^a(2,2)),\cr
[P^a(2,0),\;\tK^{b}(1,3)]&=&-\tM^{ab}(2,2)-3\ep^{ab}{}_c\tilde M^c(2,2)-\tfrac{8}{3}\eta^{ab}\tilde D(2,2),\cr
[P^a(2,0),\;K^{bc}(0,4)]&=&-4\ep^{a(b}{}_d\tK^{c)d}(1,3)-6(\eta^{a(b}\tK^{c)}(1,3)-\tfrac{1}{3}\eta^{bc}\tK^a(1,3)),\cr
[M^a(1,1),\;P^{bc}(4,0)]&=&2\ep^{a(b}{}_dP^{c)d}(4,0),\cr
[M^a(1,1),\;\tP^{bc}(3,1)&=&2\ep^{a(b}{}_d\tP^{c)d}(3,1),\cr
[M^a(1,1),\;\tP^b(3,1)]&=&\ep^{ab}{}_c\tP^c(3,1),\cr
[M^a(1,1),\;\tilde M^{bc}(2,2)]&=&2\ep^{a(b}{}_d \tilde M^{c)d}(2,2),\cr
[M^a(1,1),\;\tilde M^b(2,2)]&=&\ep^{ab}{}_c\tilde M^c(2,2),\cr
[M^a(1,1),\;\tK^{bc}(1,3)&=&2\ep^{a(b}{}_d\tK^{c)d}(1,3),\cr
[M^a(1,1),\;\tK^{b}(1,3)&=&\ep^{ab}{}_c\tK^{c}(1,3),\cr
[M^a(1,1),\;K^{bc}(0,4)&=&2\ep^{a(b}{}_dK^{c)d}(0,4),\cr
[D(1,1),\;P^{bc}(4,0)]&=&2P^{bc}(4,0),\cr
[D(1,1),\;\tP^{bc}(3,1)]&=&\tP^{bc}(3,1),\cr
[D(1,1),\;\tP^{b}(3,1)]&=&\tP^{b}(3,1),\cr
[D(1,1),\;\tK^{bc}(1,3)]&=&-\tK^{bc}(1,3),\cr
[D(1,1),\;\tK^{b}(1,3)]&=&-\tK^{b}(1,3)\cr
[D(1,1),\;K^{bc}(0,4)]&=&-2K^{bc}(0,4),\cr
[K^a(0,2),\;P^{bc}(4,0)&=&-4\ep^{a(b}{}_d\tilde P^{c)d}(3,1)+6(\eta^{a(b}\tilde P^{c)}(3,1)-\tfrac{1}{3}\eta^{bc}\tilde P^a(3,1)),\cr
[K^a(0,2),\;\tP^{bc}(3,1)]&=&3\ep^{a(b}{}_d\tilde M^{c)d}(2,2)+3(\eta^{a(b}\tilde M^{c)}(2,2)-\tfrac{1}{3}\eta^{bc}\tilde M^a(2,2)),\cr
[K^a(0,2),\;\tP^b(3,1)]&=&\tM^{ab}(2,2)-3\ep^{ab}{}_c\tilde M^c(2,2)+\tfrac{8}{3}\eta^{ab}\tilde D(2,2),\cr
[K^a(0,2),\;\tM^{bc}(2,2)]&=&-2\ep^{a(b}{}_d\tilde K^{c)d}(1,3)+(\eta^{a(b}\tilde K^{c)}(1,3)-\tfrac{1}{3}\eta^{bc}\tilde K^a(1,3)),\cr
[K^a(0,2),\;\tM^b(2,2)]&=&\tilde K^{ab}(1,3)+\tfrac{3}{2}\ep^{ab}{}_c \tilde K^c(1,3),\cr
[K^a(0,2),\;\tD(2,2)]&=&2\tilde K^a(1,3),\cr
[K^a(0,2),\;\tK^{bc}(1,3)]&=&\ep^{a(b}{}_dK^{c)d}(0,4),\cr
[K^a(0,2),\;\tK^{b}(1,3)]&=&K^{ab}(0,4).
\eeqa

\section{Some basic spin 3 equations}
\label{app:C}

\subsection{\texorpdfstring{$F_3=0$}{Spin 3 zero curvature equations}}
\label{app:C1}
\label{app:spin3eom}
The full spin 3 equations of motion in differential form notation without any gauge choice are given by (with $D=d+\tom(1,1)$ and where tr denotes the trace),
\begin{align}
  F^{ab}(4,0)&=\mr{D}e^{ab}(4,0)+e^c\wedge \te^{d(a}(3,1)\epsilon^{b)}{}_{cd}-\left(e^{(a}\wedge \te^{b)}(3,1)-\trace\right)=0\,,\\
  F^{ab}(3,1)&=\mr{D}\te^{ab}(3,1)-2e^c\wedge \tom^{d(a}(2,2)\epsilon^{b)}{}_{cd}-\left(e^{(a}\wedge \tom^{b)}(2,2)-\trace\right)\nonumber\\ &\phantom{=}-4f^c\wedge e^{d(a}(4,0)\ep^{b)}{}_{cd} =0\,,\\ 
  F^a(3,1)&=\mr{D}\te^a(3,1)-e_b\wedge \tom^{ba}(2,2)+\tfrac{3}{2}\ep^a{}_{bc}e^b
 \wedge \tom^c(2,2)-2e^a\wedge \tb(2,2)\nonumber\\ &\phantom{=}+6f_b\wedge e^{ba}(4,0)=0\,,\\
 F^{ab}(2,2)&=\mr{D}\tom^{ab}(2,2)+3e^c\wedge \tf^{d(a}(1,3)\ep_{cd}{}^{b)}+3f^c\wedge \te^{d(a}(3,1)\ep_{cd}{}^{b)}\nonumber\\ &-\left(e^{(a}\wedge \tf^{b)}(1,3)-\trace\right)+(f^{(a}\wedge \te^{b)}(3,1)-\trace)=0\,,\\
 F^{a}(2,2)&=\mr{D}\tom^{a}(2,2)-3e_b\wedge \tf^{ba}(1,3)-3\ep^a{}_{bc}e^b\wedge \tf^c(1,3)+3f_b\wedge \te^{ba}(3,1)\nonumber\\ &\phantom{=}-3\ep^a{}_{bc}f^b\wedge \te^c(3,1)=0\,,\\
 F(2,2)&=\mr{d}\tb(2,2)-\tfrac{8}{3}e^a\wedge \tf_a(1,3)+\tfrac{8}{3}f^a\wedge \te_a(3,1)=0\,,\\
   F^{ab}(1,3)&=\mr{D}\tf^{ab}(1,3)-4e^c\wedge f^{d(a}(0,4)\epsilon^{b)}{}_{cd}+\left(f^{(a}\wedge \tom^{b)}(2,2)-\trace\right)\nonumber\\ &\phantom{=}-2f^c\wedge \tom^{d(a}(2,2)\ep^{b)}{}_{cd} =0\,,\\
   F^{a}(1,3)&=\mr{D}\tf^a(1,3)-6e_b\wedge f^{ba}(0,4)+f_b\wedge \tom^{ba}(2,2)+\tfrac{3}{2}f^b\wedge \tom^c(2,2)\ep_{bc}{}^a\nonumber\\ &\phantom{=}+2f^a\wedge \tb(2,2)=0\,,\\
   F^{ab}(0,4)&=\mr{D}f^{ab}(0,4)+f^c\wedge \tf^{d(a}(1,3)\epsilon^{b)}{}_{cd}+\left(f^{(a}\wedge \tf^{b)}(1,3)-\trace\right)=0\,.
\end{align}

\subsection{\texorpdfstring{$\de A_3=(d\Lam+[A,\Lam])|_3$}{Spin 3 gauge transformations}}
\label{app:C2}

Here we give the explicit form of the gauge transformations in the spin 3 sector. They read,
using only single commutators and before implementing any gauge choices, as follows
\label{app:spin3gauge}
\input{gauges3nt.tex}

By looking at, e.g., $\de \te_{\mu}{}^a$ and the three terms involving $\Lam^{ab}(2,2), \Lam^{a}(2,2), \Lam(2,2)$ on the second line we  conclude that we can set $\te_{\mu}{}^a=0$ (at least infinitesimally) as a choice of gauge. This involves solving for these three parameters in terms of, in particular, $\mr{D}_{\mu}\Lam^a(3,1)$. Considerations like this is used in, e.g.,  section 4.

\bibliography{refs}{}
\bibliographystyle{JHEP}

\end{document}

%% file: eomspin3.tex
\begin{align} 
  \tensor{F}{_\mu_\nu^a^b}(4,0) &= \mathrm{D}_{[\mu}\tensor{e}{_{\nu]}^{a}^{b}} + \tensor{\epsilon}{_{[\mu}^{c}^{(a}}\tensor{\tilde{e}}{_{\nu]}^{b)}_{c}}, \\ 
  \tensor{F}{_\mu_\nu^a^b}(3,1) &= \mathrm{D}_{[\mu}\tensor{\tilde{e}}{_{\nu]}^{a}^{b}} - 2\tensor{\epsilon}{_{[\mu}^{c}^{(a}}\tensor{\tilde{\omega}}{_{\nu]}^{b)}_{c}} - 4\tensor{\epsilon}{^{c}^{d}^{(a}}\tensor{f}{_{c}_{[\mu}}\tensor{e}{_{\nu]}^{b)}_{d}}, \\ 
  \tensor{F}{_\mu_\nu^a}(3,1) &= -2\tensor{e}{_{[\mu}^{a}}\tensor{\tilde{b}}{_{\nu]}} + \tfrac{3}{2}\tensor{\epsilon}{_{\mu}_{\nu}^{a}}\tensor{\hat{\omega}}{} + \tensor{\tilde{\omega}}{_{[\mu}_{\nu]}^{a}} + 6\tensor{f}{^{b}_{[\mu}}\tensor{e}{_{\nu]}^{a}_{b}}, \\ 
  \tensor{F}{_\mu_\nu^a^b}(2,2) &= \mathrm{D}_{[\mu}\tensor{\tilde{\omega}}{_{\nu]}^{a}^{b}} + \left(\tensor{e}{_{[\mu}^{(a}}\tensor{\epsilon}{_{\nu]}^{b)}_{c}}\tensor{\hat{f}}{^{c}}-\trace\right) + 3\tensor{\epsilon}{_{[\mu}^{c}^{(a}}\tensor{\tilde{f}}{_{\nu]}^{b)}_{c}} + 3\tensor{\epsilon}{^{c}^{d}^{(a}}\tensor{f}{_{c}_{[\mu}}\tensor{\tilde{e}}{_{\nu]}^{b)}_{d}}, \\ 
  \tensor{F}{_\mu_\nu^a}(2,2) &= -\tensor{e}{_{[\mu}^{a}}\mathrm{D}_{\nu]}\tensor{\hat{\omega}}{} - 3\tensor{\epsilon}{_{\mu}_{\nu}^{a}}\tensor{\hat{f}}{} - 3\tensor{e}{_{[\mu}^{a}}\tensor{\hat{f}}{_{\nu]}} + 3\tensor{\tilde{f}}{_{[\mu}_{\nu]}^{a}} + 3\tensor{f}{^{b}_{[\mu}}\tensor{\tilde{e}}{_{\nu]}^{a}_{b}}, \\ 
  \tensor{F}{_\mu_\nu}(2,2) &= \mathrm{D}_{[\mu}\tensor{\tilde{b}}{_{\nu]}} - \tfrac{8}{3}\tensor{\epsilon}{_{\mu}_{\nu}_{a}}\tensor{\hat{f}}{^{a}}, \\ 
  \tensor{F}{_\mu_\nu^a^b}(1,3) &= \mathrm{D}_{[\mu}\tensor{\tilde{f}}{_{\nu]}^{a}^{b}} - 4\tensor{\epsilon}{_{[\mu}^{c}^{(a}}\tensor{f}{_{\nu]}^{b)}_{c}} - \tensor{e}{_{[\mu}^{(a}}\tensor{\hat{\omega}}{}\tensor{f}{^{b)}_{\nu]}} - 2\tensor{\epsilon}{^{c}^{d}^{(a}}\tensor{f}{_{c}_{[\mu}}\tensor{\tilde{\omega}}{_{\nu]}^{b)}_{d}}, \\ 
  \tensor{F}{_\mu_\nu^a}(1,3) &= -\tensor{e}{_{[\mu}^{a}}\mathrm{D}_{\nu]}\tensor{\hat{f}}{} - \tensor{\epsilon}{_{[\mu}^{b}^{a}}\mathrm{D}_{\nu]}\tensor{\hat{f}}{_{b}} + 2\tensor{f}{^{a}_{[\mu}}\tensor{\tilde{b}}{_{\nu]}} + 6\tensor{f}{_{[\mu}^{a}_{\nu]}} + \tfrac{3}{2}\tensor{\epsilon}{_{[\mu}^{b}^{a}}\tensor{f}{_{\nu]}_{b}}\tensor{\hat{\omega}}{} + \tensor{f}{^{b}_{[\mu}}\tensor{\tilde{\omega}}{_{\nu]}^{a}_{b}}, \\ 
  \tensor{F}{_\mu_\nu^a^b}(0,4) &= \mathrm{D}_{[\mu}\tensor{f}{_{\nu]}^{a}^{b}} - \left(\tensor{e}{_{[\mu}^{(a}}\tensor{f}{^{b)}_{\nu]}}\tensor{\hat{f}}{}-\trace\right) - \tensor{\epsilon}{_{[\mu}_{|c|}^{(a}}\tensor{f}{^{b)}_{\nu]}}\tensor{\hat{f}}{^{c}} + \tensor{\epsilon}{^{c}^{d}^{(a}}\tensor{f}{_{c}_{[\mu}}\tensor{\tilde{f}}{_{\nu]}^{b)}_{d}}.
\end{align}

%% file: spin3solutions.tex
\begin{align} 
\tensor{\tilde{e}}{_{\mu}^{a}^{b}}&=\tensor{\epsilon}{_{\mu}^{c}^{d}}\mathrm{D}_{c}\tensor{e}{_{d}^{a}^{b}} - 2\tensor{\epsilon}{^{c}^{d}^{(a}}\mathrm{D}_{c}\tensor{e}{_{d}^{b)}_{\mu}}-\trace,\\ 
\tensor{\tilde{\omega}}{_{\mu}^{a}^{b}}&=-\frac{1}{2}\tensor{\epsilon}{_{\mu}^{c}^{d}}\mathrm{D}_{c}\tensor{\tilde{e}}{_{d}^{a}^{b}} + \tensor{\epsilon}{^{c}^{d}^{(a}}\mathrm{D}_{c}\tensor{\tilde{e}}{_{d}^{b)}_{\mu}} - 2\tensor{f}{^{a}^{b}}\tensor{e}{_{c}_{\mu}^{c}} + 2\tensor{f}{^{c}_{\mu}}\tensor{e}{_{c}^{a}^{b}} - 2\tensor{f}{_{c}^{c}}\tensor{e}{_{\mu}^{a}^{b}} + 2\tensor{f}{^{c}^{(a}}\tensor{e}{_{\mu}^{b)}_{c}}\nonumber\\ 
  &\phantom{=}-2\tensor{f}{^{c}_{\mu}}\tensor{e}{^{(a}^{b)}_{c}} + 2\tensor{f}{^{c}^{(a}}\tensor{e}{^{b)}_{c}_{\mu}}-\trace,\label{tesolution}\\ 
\tensor{\tilde{b}}{_{\mu}}&=-\frac{1}{9}\mathrm{D}^{a}\mathrm{D}_{a}\tensor{e}{_{b}_{\mu}^{b}} - \frac{1}{18}\mathrm{D}^{a}\mathrm{D}_{\mu}\tensor{e}{_{b}_{a}^{b}} + \frac{1}{9}\mathrm{D}^{a}\mathrm{D}^{b}\tensor{e}{_{a}_{b}_{\mu}} + \frac{1}{18}\mathrm{D}^{a}\mathrm{D}^{b}\tensor{e}{_{\mu}_{a}_{b}} + \frac{1}{3}\tensor{f}{^{a}^{b}}\tensor{e}{_{a}_{b}_{\mu}} - \frac{4}{3}\tensor{f}{_{\mu}^{a}}\tensor{e}{_{b}_{a}^{b}}\nonumber\\ 
  &\phantom{=}-\frac{1}{3}\tensor{f}{_{a}^{a}}\tensor{e}{_{b}_{\mu}^{b}} + \frac{4}{3}\tensor{f}{^{a}^{b}}\tensor{e}{_{\mu}_{a}_{b}},\\ 
\tensor{\hat{\omega}}{}&=\frac{2}{3}\tensor{\epsilon}{^{a}^{b}^{c}}\tensor{f}{_{a}^{d}}\tensor{e}{_{b}_{c}_{d}},\\ 
\tensor{\hat{f}}{_{\mu}}&=-\frac{1}{3}\mathrm{D}_{\mu}\tensor{\hat{\omega}}{} + \frac{1}{2}\tensor{\tilde{f}}{_{a}_{\mu}^{a}} - \frac{1}{2}\tensor{f}{_{\mu}^{a}}\tensor{\tilde{e}}{_{b}_{a}^{b}} + \frac{1}{2}\tensor{f}{^{a}^{b}}\tensor{\tilde{e}}{_{\mu}_{a}_{b}},\\ 
\tensor{\hat{f}}{}&=-\frac{2}{3}\tensor{f}{^{a}^{b}}\mathrm{D}_{[a}\tensor{e}{_{c]}_{b}^{c}},\\
\tensor{\tilde{f}}{_{\mu}^{a}^{b}}&=\frac{1}{3}\tensor{\epsilon}{_{\mu}^{c}^{d}}\mathrm{D}_{c}\tensor{\tilde{\omega}}{_{d}^{a}^{b}} - \frac{1}{48}\tensor{e}{_{\mu}^{(a}}\tensor{\epsilon}{^{|c}^{d}^{m|}}\mathrm{D}_{c}\tensor{\tilde{\omega}}{_{d}^{b)}_{m}} - \frac{2}{3}\tensor{\epsilon}{^{c}^{d}^{(a}}\mathrm{D}_{c}\tensor{\tilde{\omega}}{_{d}^{b)}_{\mu}} + \frac{1}{16}\tensor{e}{_{\mu}^{(a}}\mathrm{D}^{b)}\tensor{\hat{\omega}}{} - \tensor{f}{^{a}^{b}}\tensor{\tilde{e}}{_{c}_{\mu}^{c}} + \tensor{f}{^{c}_{\mu}}\tensor{\tilde{e}}{_{c}^{a}^{b}}\nonumber\\ 
  &\phantom{=}+\frac{1}{16}\tensor{e}{_{\mu}^{(a}}\tensor{f}{^{b)}^{c}}\tensor{\tilde{e}}{_{d}_{c}^{d}} - \frac{1}{16}\tensor{e}{_{\mu}^{(a}}\tensor{f}{^{|c}^{d|}}\tensor{\tilde{e}}{_{d}^{b)}_{c}} + \frac{1}{16}\tensor{e}{_{\mu}^{(a}}\tensor{f}{_{c}^{|c|}}\tensor{\tilde{e}}{_{d}^{b)}^{d}} - \tensor{f}{_{c}^{c}}\tensor{\tilde{e}}{_{\mu}^{a}^{b}} + \tensor{f}{^{c}^{(a}}\tensor{\tilde{e}}{_{\mu}^{b)}_{c}} - \tensor{f}{^{c}_{\mu}}\tensor{\tilde{e}}{^{(a}^{b)}_{c}}\nonumber\\ 
  &\phantom{=}-\frac{1}{16}\tensor{e}{_{\mu}^{(a}}\tensor{f}{^{|c}^{d|}}\tensor{\tilde{e}}{^{b)}_{c}_{d}} + \tensor{f}{^{c}^{(a}}\tensor{\tilde{e}}{^{b)}_{c}_{\mu}}-\trace,\\ 
\tensor{f}{_{\mu}^{a}^{b}}&=-\frac{1}{4}\tensor{\epsilon}{_{\mu}^{c}^{d}}\mathrm{D}_{c}\tensor{\tilde{f}}{_{d}^{a}^{b}} + \frac{1}{2}\tensor{\epsilon}{^{c}^{d}^{(a}}\mathrm{D}_{c}\tensor{\tilde{f}}{_{d}^{b)}_{\mu}} - \frac{1}{2}\tensor{\epsilon}{_{\mu}^{c}^{(a}}\tensor{\hat{\omega}}{}\tensor{f}{^{b)}_{c}} - \frac{1}{2}\tensor{f}{^{a}^{b}}\tensor{\tilde{\omega}}{_{c}_{\mu}^{c}} + \frac{1}{2}\tensor{f}{^{c}_{\mu}}\tensor{\tilde{\omega}}{_{c}^{a}^{b}} - \frac{1}{2}\tensor{f}{_{c}^{c}}\tensor{\tilde{\omega}}{_{\mu}^{a}^{b}}\nonumber\\ 
  &\phantom{=}+\frac{1}{2}\tensor{f}{^{c}^{(a}}\tensor{\tilde{\omega}}{_{\mu}^{b)}_{c}} - \frac{1}{2}\tensor{f}{^{c}_{\mu}}\tensor{\tilde{\omega}}{^{(a}^{b)}_{c}} + \frac{1}{2}\tensor{f}{^{c}^{(a}}\tensor{\tilde{\omega}}{^{b)}_{c}_{\mu}}-\trace.\label{fsolution}\end{align}

%% file: gauges3nt.tex
\begin{align} 
  \delta\underset{(4,0)}{\tensor{e}{_{\mu}^{a}^{b}}} &= \mathrm{D}_{\mu}\underset{(4,0)}{\tensor{\Lambda}{^{a}^{b}}} + 2\tensor{\epsilon}{^{(a}^{|c}^{d|}}\tensor{e}{_{\mu}^{b)}_{c}}\underset{(1,1)}{\tensor{\Lambda}{_{d}}} - 2\tensor{e}{_{\mu}^{a}^{b}}\underset{(1,1)}{\tensor{\Lambda}{}} + \tensor{\epsilon}{^{(a}^{|c}^{d|}}\tensor{\tilde{e}}{_{\mu}^{b)}_{c}}\underset{(2,0)}{\tensor{\Lambda}{_{d}}} + \tensor{\tilde{e}}{_{\mu}^{(a}}\underset{(2,0)}{\tensor{\Lambda}{^{b)}}}\nonumber\\ 
  &\phantom{=}-\tensor{\epsilon}{^{(a}^{|d|}_{\mu}}\underset{(3,1)}{\tensor{\Lambda}{^{b)}_{d}}} - \tensor{e}{_{\mu}^{(a}}\underset{(3,1)}{\tensor{\Lambda}{^{b)}}} + 2\tensor{b}{_{\mu}}\underset{(4,0)}{\tensor{\Lambda}{^{a}^{b}}}, \\ 
  \delta\underset{(3,1)}{\tensor{\tilde{e}}{_{\mu}^{a}}} &= \mathrm{D}_{\mu}\underset{(3,1)}{\tensor{\Lambda}{^{a}}} - 6\tensor{e}{_{\mu}^{a}^{b}}\underset{(0,2)}{\tensor{\Lambda}{_{b}}} + \tensor{\epsilon}{^{a}^{b}^{c}}\tensor{\tilde{e}}{_{\mu}_{b}}\underset{(1,1)}{\tensor{\Lambda}{_{c}}} - \tensor{\tilde{e}}{_{\mu}^{a}}\underset{(1,1)}{\tensor{\Lambda}{}} + \tensor{\tilde{\omega}}{_{\mu}^{a}^{b}}\underset{(2,0)}{\tensor{\Lambda}{_{b}}} + \frac{3}{2}\tensor{\epsilon}{^{a}^{b}^{c}}\tensor{\tilde{\omega}}{_{\mu}_{b}}\underset{(2,0)}{\tensor{\Lambda}{_{c}}}\nonumber\\ 
  &\phantom{=}+2\tensor{\tilde{b}}{_{\mu}}\underset{(2,0)}{\tensor{\Lambda}{^{a}}} - \underset{(2,2)}{\tensor{\Lambda}{^{a}_{\mu}}} - \frac{3}{2}\tensor{\epsilon}{^{a}^{c}_{\mu}}\underset{(2,2)}{\tensor{\Lambda}{_{c}}} - 2\tensor{e}{_{\mu}^{a}}\underset{(2,2)}{\tensor{\Lambda}{}} + \tensor{b}{_{\mu}}\underset{(3,1)}{\tensor{\Lambda}{^{a}}} + 6\tensor{f}{_{\mu}^{b}}\underset{(4,0)}{\tensor{\Lambda}{^{a}_{b}}}, \\ 
  \delta\underset{(3,1)}{\tensor{\tilde{e}}{_{\mu}^{a}^{b}}} &= \mathrm{D}_{\mu}\underset{(3,1)}{\tensor{\Lambda}{^{a}^{b}}} - 4\tensor{\epsilon}{^{(a}^{|c}^{d|}}\tensor{e}{_{\mu}^{b)}_{c}}\underset{(0,2)}{\tensor{\Lambda}{_{d}}} + 2\tensor{\epsilon}{^{(a}^{|c}^{d|}}\tensor{\tilde{e}}{_{\mu}^{b)}_{c}}\underset{(1,1)}{\tensor{\Lambda}{_{d}}} - \tensor{\tilde{e}}{_{\mu}^{a}^{b}}\underset{(1,1)}{\tensor{\Lambda}{}} - 2\tensor{\epsilon}{^{(a}^{|c}^{d|}}\tensor{\tilde{\omega}}{_{\mu}^{b)}_{c}}\underset{(2,0)}{\tensor{\Lambda}{_{d}}}\nonumber\\ 
  &\phantom{=}+\tensor{\tilde{\omega}}{_{\mu}^{(a}}\underset{(2,0)}{\tensor{\Lambda}{^{b)}}} + 2\tensor{\epsilon}{^{(a}^{|d|}_{\mu}}\underset{(2,2)}{\tensor{\Lambda}{^{b)}_{d}}} - \tensor{e}{_{\mu}^{(a}}\underset{(2,2)}{\tensor{\Lambda}{^{b)}}} + \tensor{b}{_{\mu}}\underset{(3,1)}{\tensor{\Lambda}{^{a}^{b}}} - 4\tensor{\epsilon}{^{(a}^{|c}^{d|}}\tensor{f}{_{\mu}_{c}}\underset{(4,0)}{\tensor{\Lambda}{^{b)}_{d}}}, \\ 
  \delta\underset{(2,2)}{\tensor{\tilde{b}}{_{\mu}}} &= \mathrm{D}_{\mu}\underset{(2,2)}{\tensor{\Lambda}{}} - \frac{8}{3}\tensor{\tilde{e}}{_{\mu}^{a}}\underset{(0,2)}{\tensor{\Lambda}{_{a}}} - \frac{8}{3}\underset{(1,3)}{\tensor{\Lambda}{_{\mu}}} + \frac{8}{3}\tensor{\tilde{f}}{_{\mu}^{a}}\underset{(2,0)}{\tensor{\Lambda}{_{a}}} + \frac{8}{3}\tensor{f}{_{\mu}^{a}}\underset{(3,1)}{\tensor{\Lambda}{_{a}}}, \\ 
  \delta\underset{(2,2)}{\tensor{\tilde{\omega}}{_{\mu}^{a}}} &= \mathrm{D}_{\mu}\underset{(2,2)}{\tensor{\Lambda}{^{a}}} - 3\tensor{\tilde{e}}{_{\mu}^{a}^{b}}\underset{(0,2)}{\tensor{\Lambda}{_{b}}} - 3\tensor{\epsilon}{^{a}^{b}^{c}}\tensor{\tilde{e}}{_{\mu}_{b}}\underset{(0,2)}{\tensor{\Lambda}{_{c}}} + \tensor{\epsilon}{^{a}^{b}^{c}}\tensor{\tilde{\omega}}{_{\mu}_{b}}\underset{(1,1)}{\tensor{\Lambda}{_{c}}} - 3\underset{(1,3)}{\tensor{\Lambda}{^{a}_{\mu}}} + 3\tensor{\epsilon}{^{a}^{c}_{\mu}}\underset{(1,3)}{\tensor{\Lambda}{_{c}}}\nonumber\\ 
  &\phantom{=}+3\tensor{\tilde{f}}{_{\mu}^{a}^{b}}\underset{(2,0)}{\tensor{\Lambda}{_{b}}} - 3\tensor{\epsilon}{^{a}^{b}^{c}}\tensor{\tilde{f}}{_{\mu}_{b}}\underset{(2,0)}{\tensor{\Lambda}{_{c}}} + 3\tensor{f}{_{\mu}^{b}}\underset{(3,1)}{\tensor{\Lambda}{^{a}_{b}}} - 3\tensor{\epsilon}{^{a}^{b}^{c}}\tensor{f}{_{\mu}_{b}}\underset{(3,1)}{\tensor{\Lambda}{_{c}}}, \\ 
  \delta\underset{(2,2)}{\tensor{\tilde{\omega}}{_{\mu}^{a}^{b}}} &= \mathrm{D}_{\mu}\underset{(2,2)}{\tensor{\Lambda}{^{a}^{b}}} + 3\tensor{\epsilon}{^{(a}^{|c}^{d|}}\tensor{\tilde{e}}{_{\mu}^{b)}_{c}}\underset{(0,2)}{\tensor{\Lambda}{_{d}}} - \tensor{\tilde{e}}{_{\mu}^{(a}}\underset{(0,2)}{\tensor{\Lambda}{^{b)}}} + 2\tensor{\epsilon}{^{(a}^{|c}^{d|}}\tensor{\tilde{\omega}}{_{\mu}^{b)}_{c}}\underset{(1,1)}{\tensor{\Lambda}{_{d}}} - 3\tensor{\epsilon}{^{(a}^{|d|}_{\mu}}\underset{(1,3)}{\tensor{\Lambda}{^{b)}_{d}}}\nonumber\\ 
  &\phantom{=}-\tensor{e}{_{\mu}^{(a}}\underset{(1,3)}{\tensor{\Lambda}{^{b)}}} + 3\tensor{\epsilon}{^{(a}^{|c}^{d|}}\tensor{\tilde{f}}{_{\mu}^{b)}_{c}}\underset{(2,0)}{\tensor{\Lambda}{_{d}}} + \tensor{\tilde{f}}{_{\mu}^{(a}}\underset{(2,0)}{\tensor{\Lambda}{^{b)}}} + 3\tensor{\epsilon}{^{(a}^{|c}^{d|}}\tensor{f}{_{\mu}_{c}}\underset{(3,1)}{\tensor{\Lambda}{^{b)}_{d}}} + \tensor{f}{_{\mu}^{(a}}\underset{(3,1)}{\tensor{\Lambda}{^{b)}}}, \\ 
  \delta\underset{(1,3)}{\tensor{\tilde{f}}{_{\mu}^{a}}} &= \mathrm{D}_{\mu}\underset{(1,3)}{\tensor{\Lambda}{^{a}}} - \tensor{\tilde{\omega}}{_{\mu}^{a}^{b}}\underset{(0,2)}{\tensor{\Lambda}{_{b}}} + \frac{3}{2}\tensor{\epsilon}{^{a}^{b}^{c}}\tensor{\tilde{\omega}}{_{\mu}_{b}}\underset{(0,2)}{\tensor{\Lambda}{_{c}}} - 2\tensor{\tilde{b}}{_{\mu}}\underset{(0,2)}{\tensor{\Lambda}{^{a}}} - 6\underset{(0,4)}{\tensor{\Lambda}{^{a}_{\mu}}} + \tensor{\epsilon}{^{a}^{b}^{c}}\tensor{\tilde{f}}{_{\mu}_{b}}\underset{(1,1)}{\tensor{\Lambda}{_{c}}} + \tensor{\tilde{f}}{_{\mu}^{a}}\underset{(1,1)}{\tensor{\Lambda}{}}\nonumber\\ 
  &\phantom{=}-\tensor{b}{_{\mu}}\underset{(1,3)}{\tensor{\Lambda}{^{a}}} + 6\tensor{f}{_{\mu}^{a}^{b}}\underset{(2,0)}{\tensor{\Lambda}{_{b}}} + \tensor{f}{_{\mu}^{b}}\underset{(2,2)}{\tensor{\Lambda}{^{a}_{b}}} + \frac{3}{2}\tensor{\epsilon}{^{a}^{b}^{c}}\tensor{f}{_{\mu}_{b}}\underset{(2,2)}{\tensor{\Lambda}{_{c}}} + 2\tensor{f}{_{\mu}^{a}}\underset{(2,2)}{\tensor{\Lambda}{}}, \\ 
  \delta\underset{(1,3)}{\tensor{\tilde{f}}{_{\mu}^{a}^{b}}} &= \mathrm{D}_{\mu}\underset{(1,3)}{\tensor{\Lambda}{^{a}^{b}}} - 2\tensor{\epsilon}{^{(a}^{|c}^{d|}}\tensor{\tilde{\omega}}{_{\mu}^{b)}_{c}}\underset{(0,2)}{\tensor{\Lambda}{_{d}}} - \tensor{\tilde{\omega}}{_{\mu}^{(a}}\underset{(0,2)}{\tensor{\Lambda}{^{b)}}} + 4\tensor{\epsilon}{^{(a}^{|d|}_{\mu}}\underset{(0,4)}{\tensor{\Lambda}{^{b)}_{d}}} + 2\tensor{\epsilon}{^{(a}^{|c}^{d|}}\tensor{\tilde{f}}{_{\mu}^{b)}_{c}}\underset{(1,1)}{\tensor{\Lambda}{_{d}}}\nonumber\\ 
  &\phantom{=}+\tensor{\tilde{f}}{_{\mu}^{a}^{b}}\underset{(1,1)}{\tensor{\Lambda}{}} - \tensor{b}{_{\mu}}\underset{(1,3)}{\tensor{\Lambda}{^{a}^{b}}} - 4\tensor{\epsilon}{^{(a}^{|c}^{d|}}\tensor{f}{_{\mu}^{b)}_{c}}\underset{(2,0)}{\tensor{\Lambda}{_{d}}} - 2\tensor{\epsilon}{^{(a}^{|c}^{d|}}\tensor{f}{_{\mu}_{c}}\underset{(2,2)}{\tensor{\Lambda}{^{b)}_{d}}} + \tensor{f}{_{\mu}^{(a}}\underset{(2,2)}{\tensor{\Lambda}{^{b)}}}, \\ 
  \delta\underset{(0,4)}{\tensor{f}{_{\mu}^{a}^{b}}} &= \mathrm{D}_{\mu}\underset{(0,4)}{\tensor{\Lambda}{^{a}^{b}}} + \tensor{\epsilon}{^{(a}^{|c}^{d|}}\tensor{\tilde{f}}{_{\mu}^{b)}_{c}}\underset{(0,2)}{\tensor{\Lambda}{_{d}}} - \tensor{\tilde{f}}{_{\mu}^{(a}}\underset{(0,2)}{\tensor{\Lambda}{^{b)}}} - 2\tensor{b}{_{\mu}}\underset{(0,4)}{\tensor{\Lambda}{^{a}^{b}}} + 2\tensor{\epsilon}{^{(a}^{|c}^{d|}}\tensor{f}{_{\mu}^{b)}_{c}}\underset{(1,1)}{\tensor{\Lambda}{_{d}}}\nonumber\\ 
  &\phantom{=}+2\tensor{f}{_{\mu}^{a}^{b}}\underset{(1,1)}{\tensor{\Lambda}{}} + \tensor{\epsilon}{^{(a}^{|c}^{d|}}\tensor{f}{_{\mu}_{c}}\underset{(1,3)}{\tensor{\Lambda}{^{b)}_{d}}} + \tensor{f}{_{\mu}^{(a}}\underset{(1,3)}{\tensor{\Lambda}{^{b)}}}.
\end{align}